\documentclass[showpacs,amsmath,amssymb,graphicx,12pt]{revtex4}
\usepackage{graphicx}
\usepackage{epsfig}
\usepackage{multirow}
\usepackage{subfigure}
\usepackage{dcolumn}
\usepackage{bm}

\begin{document}

\title{A possible $NN^{*}(1440)$ quasi-molecular state}

\author{Lu Zhao$^{1,2}$, Peng-Nian Shen$^{3,1,4}$, Ying-Jie Zhang$^{5,4}$, and Bing-Song
Zou$^{6,1,2,4}$ }

\affiliation{1. Institute of High Energy Physics, CAS, Beijing
100049, China
\\
2. University of Chinese Academy of Sciences, Beijing 100049, China
\\
3. College of Physics and Technology, Guangxi Normal University,
Guilin  541004, China
\\4. Theoretical Physics Center for Science Facilities, CAS,
Beijing 100049, China
\\5. Physics Department, Hebei University,  Baoding 071002, China
\\6. State Key Laboratory of Theoretical Physics, Institute of Theoretical Physics, CAS, Beijing
100190, China }

\begin{abstract}
Inspired by the recent observation of a narrow resonance-like
structure around 2360 MeV in the $pn\to d\pi^0\pi^0$ cross section,
the possibility of forming a $NN^{*}$(1440) quasi-molecular state is
investigated by using a meson exchange model in which the $\pi$,
$\sigma$, $\rho$ and $\omega$ exchanges in t- and u-channels are
considered. By adopting the coupling constants extracted from the
relevant $NN$ scattering and $N^*(1440)$ decay data, it is found
that a deuteron-like  quasi-molecular state of $NN^{*}(1440)$ with a
binding energy in the range of $2\sim 67MeV$ can be formed.
Therefore, it is speculated that the observed structure around 2360
MeV might be or may have a large component of the $NN^*$(1440)
quasi-molecular state.
\end{abstract}

 \maketitle{}

\section{INTRODUCTION}

The ABC effect, which shows an unexpected peak in the invariant
$\pi$$\pi$-mass spectrum of the double pionic $D$-$p$ fusion to
$^{3}He$ process, was observed by Abashian, Booth and Crowe long
ago\cite{ABC1}. Later on, in the hadronic experiment, such a
phenomenon was also observed in the other processes, for instance
$n$ + $p$ $\rightarrow$ $D$($\pi$$\pi$)~\cite{ABC2,ABC3} and $D$ +
$D$ $\rightarrow$ $^{4}He$($\pi$$\pi$)~\cite{ABC4,ABC5}. The
2$\pi$-production mechanism used to be attributed to the effect of
the double $\Delta$ excitation. But it was found that the obtained
enhancement in the low $\pi$$\pi$ mass region in the exclusive
$\pi^{0}$$\pi^{0}$ channel is much smaller than the data
value~\cite{ABC2delta1,ABC2delta2}, which implies that some
important physics is missing in such a mechanism. Thus, a sequential
mechanism was proposed to explain the deuteron spectrum of the $n$ +
$p$ $\rightarrow$ $D$($\pi$$\pi$) process, namely the nucleon can
firstly be excited into Roper resonance $N^{*}$(1440), and then the
resonance decays into $N$$\pi$$\pi$ and $\Delta$$\pi$
subsequently~\cite{E.oset1998}.

On the other hand, the deuteron is a unique B=2 state which has been
confirmed in the experiment at this moment, where B stands for the
baryon number. Actually, after three decades of experimental
efforts, we have never found any decisive sign for the existence of
a B=2 state other than the deuteron. Recently, the CELSIUS-WASA
Collaboration observed a narrow resonance-like structure around 2360
MeV in the data set of the $pn\to d\pi^0\pi^0$ total cross
section~\cite{Bashkanov:2008ih}. Since such an energy is fairly
close to the $NN^*(1440)$ threshold, and $N^*(1440)$ has the same
quantum numbers as the $N\sigma$ system and would couple to
$N\sigma$ strongly, this phenomenon awakens our interest in
exploring the possible existence of a $NN^*(1440)$ quasi-molecular
state, which is an analogue of the deuteron.

In order to carry out such a study, the hadron-hadron interaction
should be known before hand. The hadron-hadron interaction used to
be studied in the hadron degrees of freedom. Since the QCD theory
emerged in the 1970s, such an interaction has further been
investigated in the quark degrees of freedom, where the
hadron-hadron system is considered as a multi-quark cluster system.
One of the models in this category, which can basically explain the
experiment data, is called constituent quark model (CQM). In this
model, the quark-quark interaction can be described by the
one-gluon-exchange (OGE) potential in the short-range, the scalar-
and pseudoscalar-meson-exchange potentials in the medium-range
interaction, and the phenomenological confinement potential in the
long-range.

However, because of the property of the non-Abelian group, the
non-perturbative effect of QCD still cannot be taken into account
accurately up to now. For simplicity and efficiency, the method in
the framework of the hadron degrees of freedom is still widely used
in the hadron physics investigation. In this framework, the hadron
is assumed as a point-like particle, and the hadron-hadron
interaction is supposed to be originated from exchanges of various
mesons whose masses and coupling forms with various hadrons are
different. A typical theoretical model is the Bonn meson-exchange
model which has intensively been studied over years. By employing
this model, a large amount of nucleon-nucleon interaction data, such
as the deuteron property, scattering phase shifts and reaction cross
sections, can be well described~\cite{R.machleidt1987}. This model
has also been adopted to study the baryon-baryon interaction in the
coupled channel approach, although the result was not accurate
enough due to too much approximation~\cite{F.osterfeld1997}.
Recently, a simplified one-pion-exchange model was used to study
systematically the deuteron-like meson-meson bound state, called
deuson, some possible $S$-wave molecular-like states in the
meson-meson interaction were predicted~\cite{X.liu,F.E.close}. The
interaction of $\Lambda(1405)$ with a nucleon was studied from the
viewpoint of chiral dynamics and a possible quasi-bound state was
found~\cite{Oka}.

In this paper, the Bonn meson-exchange model is extended to the
$NN^{*}$(1440) system so that the binding property of the system can
be studied. The paper is organized in the following way: The
derivation of the non-relativistic $N$-$N^{*}$(1440)potential is
briefly introduced in section II, and the model parameters are
extracted from the available data in section III. In section IV, the
numerical result and discussion are given. Finally, the summary is
presented in section V.

\section{Brief formulism}

\begin{figure}[ht]
  \begin{center}
  \rotatebox{0}{\includegraphics*[width=0.3\textwidth]{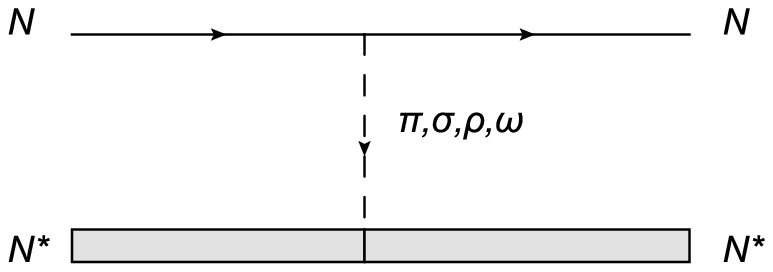}}
    \rotatebox{0}{\includegraphics*[width=0.3\textwidth]{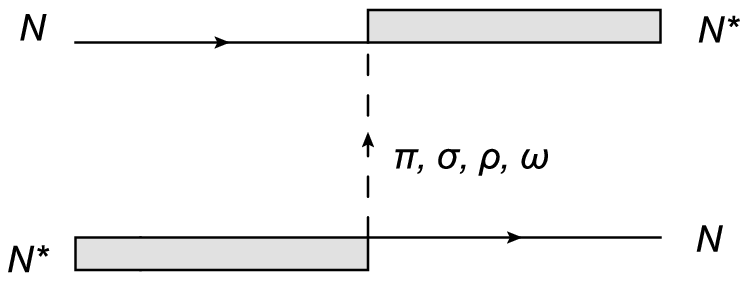}}
    \caption{ The t-channel (left) and u-channel (right) Feynman diagrams of the $N$-$N^*(1440)$ interaction.
}
    \label{Feynman}
  \end{center}
\end{figure}

The basic Feynman diagrams of the $N$-$N^*(1440)$ interaction are shown in Fig.\ref{Feynman}.
In the Bonn meson-exchange model, there are six mesons ( $\pi$, $\sigma$, $\rho$, $\omega$, $\delta$ and $\eta$) to be considered. Recently, by fitting the experimental data, one showed that only four of them, namely $\pi$, $\sigma$, $\rho$, and $\omega$, have relatively large coupling with $NN^{*}$(1440)~\cite{xie}. Clearly, the $\pi$- and $\sigma$-meson exchanges would be responsible the the long-range ($r$$\geqslant$2 fm) and the intermediate-range(1 fm$\leqslant$$r$$\leqslant$2 fm) interactions, respectively, and the $\rho$- and $\omega$-meson exchanges would provide the core-range($r$$\leqslant$1 fm) interaction.

The effective Lagrangians of the meson-$N$-$N^{*}$(1440) interaction for various mesons read
\begin{eqnarray}
\mathcal{L}_{\pi NN^*} \!\! &=&\!\! -ig_{\pi
NN^*}\bar u_N\gamma _5 u_{N^*} \vec\tau\cdot\vec\pi+ h.c. ,\\
\mathcal{L}_{\sigma NN^*}  \!\! &=&\!\!   g_{\sigma NN^*} \bar
u_N u_{N^*} \sigma + h.c. , \\
\mathcal{L}_{\rho NN^*} \!\! &=&\!\!- g_{\rho
NN^*}\bar u_N \gamma^{\mu} u_{N^*} \vec\tau\cdot\vec\rho_\mu \! - \! \frac{f_{\rho
NN^*}}{2m_{N^*}\!+\!2m_N}  \bar u_N\sigma^{\mu\nu} u_{N^*} \vec\tau\cdot
( \partial_\mu\vec\rho_\nu - \partial_\nu\vec\rho_\mu)+ h.c. ,\\
\mathcal{L}_{\omega NN^*} \!\! &=&\!\!- g_{\omega
NN^*}\bar u_N \gamma^{\mu} u_{N^*} \omega_\mu \! - \! \frac{f_{\omega
NN^*}}{2m_{N^*}\!+\!2m_N}  \bar u_N\sigma^{\mu\nu} u_{N^*}
( \partial_\mu\omega_\nu - \partial_\nu\omega_\mu)+ h.c. ,
\end{eqnarray}
The forms of the effective Lagrangians of the meson-$N$-$N$ or
meson-$N^*$-$N^*$  interaction are the same as those of the
meson-$N$-$N^*(1440)$ interaction, except the spinor.

\begin{figure}[ht]
  \begin{center}
    \rotatebox{0}{\includegraphics*[width=0.7\textwidth]{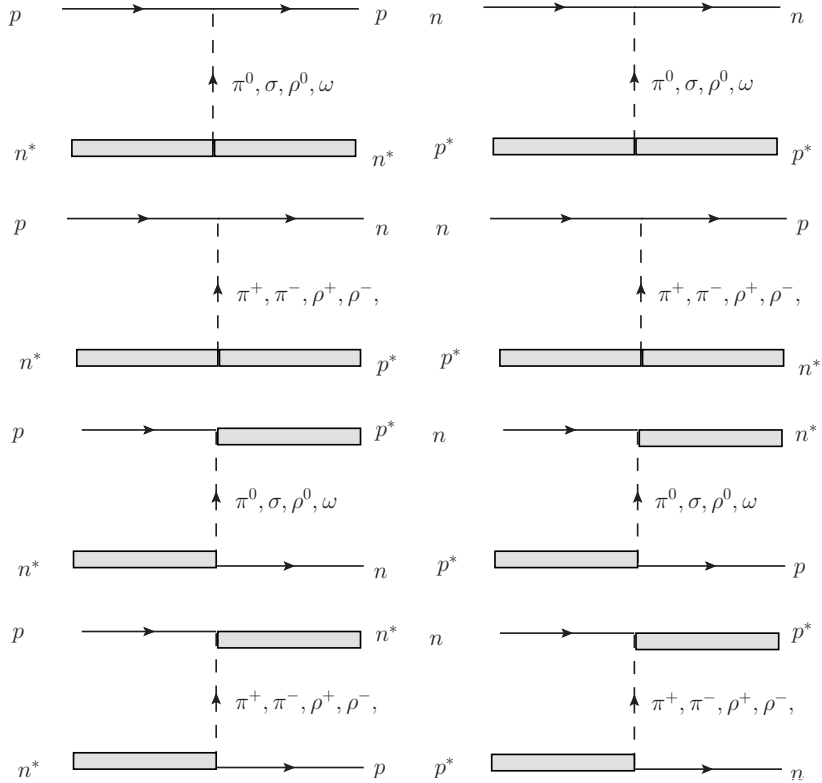}}
    \caption{8 non-equivalence Feynman diagrams in $\langle \Psi|\Psi\rangle$.\label{non-equivalence}}
  \end{center}
\end{figure}

Now, we consider a $NN^*$(1440) system whose quantum numbers are the
same as  those of the deuteron, namely $I=0$, $J^P=1^+$. Since $N^*$
is also an isospin doublet, one can write the $|NN^*\rangle$ or
$|N^*N\rangle $ state as
\begin{eqnarray}
|NN^*\rangle &=& -\frac{\sqrt{2}}{2}|np^*\rangle + \frac{\sqrt{2}}{2}|pn^*\rangle,\\
|N^*N\rangle &=& -\frac{\sqrt{2}}{2}|n^*p\rangle + \frac{\sqrt{2}}{2}|p^*n\rangle,
\end{eqnarray}
and its overlap as
\begin{eqnarray}
\langle NN^*|NN^*\rangle &=& \frac{1}{2}\langle np^*|np^*\rangle + \frac{1}{2}\langle pn^*|pn^*\rangle
-\frac{1}{2}\langle pn^*|np^*\rangle-\frac{1}{2}\langle np^*|pn^*\rangle,\nonumber\\
\langle NN^*|N^*N\rangle &=& \frac{1}{2}\langle np^*|n^*p\rangle - \frac{1}{2}\langle np^*|p^*n\rangle
-\frac{1}{2}\langle pn^*|n^*p\rangle + \frac{1}{2}\langle pn^*|p^*n\rangle,\nonumber\\
\langle N^*N|N^*N\rangle &=& \frac{1}{2}\langle n^*p|n^*p\rangle + \frac{1}{2}\langle p^*n|p^*n\rangle
-\frac{1}{2}\langle p^*n|n^*p\rangle-\frac{1}{2}\langle n^*p|p^*n\rangle,\nonumber\\
\langle N^*N|NN^*\rangle &=& \frac{1}{2}\langle n^*p|np^*\rangle - \frac{1}{2}\langle n^*p|pn^*\rangle
-\frac{1}{2}\langle p^*n|np^*\rangle+\frac{1}{2}\langle p^*n|pn^*\rangle.
\end{eqnarray}
Consequently, one can further write the wave function of the $NN^*$(1440) system as
\begin{eqnarray}
|\Psi\rangle &=& \frac{1}{\sqrt{2}}(~|NN^*\rangle + |N^*N\rangle~),
\end{eqnarray}
and the normalization as
\begin{eqnarray}
\langle \Psi|\Psi\rangle &=& \frac{1}{2}[~\langle NN^*|NN^*\rangle + \langle NN^*|N^*N\rangle
+\langle N^*N|N^*N\rangle + \langle N^*N|NN^*\rangle~]. \label{eq:norm}
\end{eqnarray}
Apparently, each term in the above equation contains four
non-equivalent Feynman diagrams.  All together, Eq.(\ref{eq:norm})
contains eight non-equivalent Feynman diagrams, as shown in
Fig.\ref{non-equivalence}. Among them, the first four are
$u$-channel diagrams and the rest are $t$-channel diagrams.

Next, we derive the $N$-$N^*$ potential for the above mentioned
diagrams  according to the quantum scattering theory. The
relativistic S-matrix for a scattering process reads
\begin{equation}
\langle f | S | i \rangle
= \delta_{fi} + (2\pi)^4\delta^4(p_f-p_i) i M_{fi},
\end{equation}
where $M_{fi}$ denotes the scattering amplitude from the $i$ state
to  the $f$ state, or the Lorentz-invariant matrix element between
the $i$ state and the $f$ state. On the other hand, by applying Bonn
approximation on the Lippmann-Schwinger equation, the S-matrix reads
\begin{equation}
\langle f | S | i \rangle = \delta_{fi} - 2\pi \delta(E_f-E_i) i V_{fi}
\end{equation}
with $V_{fi}$ being the interaction potential. Considering the
different  normalization conventions used for the scattering
amplitude $M_{fi}$ and for the $T$-matrix $T_{fi}$, and consequently
$V_{fi}$, we have
\begin{equation}
V_{fi}=-\frac{M_{fi}}{\sqrt{ \mathop\prod\limits_{f}\frac{{P_f}^0}{m_f} \mathop\prod\limits_{i} \frac{{P_i}^0}{m_i}}}
\approx -M_{fi},\label{eq:vq}
\end{equation}
where $P_{f(i)}$ denotes the four momentum of the final (initial) state.

Let us use, in the center mass system, $P_1(E_1,\vec{p})$ and
$P_2(E_2,-\vec{p})$  to represent the four momenta of the initial
particles, $P_3(E_3,\vec{p'})$ and $P_4(E_4,-\vec{p'})$ to denote
the four momenta of the final particles, respectively. And then,
\begin{equation}
q=P_3-P_1=(E_3-E_1,\vec{p'}-\vec{p})=(E_2-E_4,\vec{q})
\end{equation}
should be the transferred four momentum or the four momentum of the
meson propagator. For convenience, we always use
\begin{equation}
\vec{q}=\vec{p'}-\vec{p}
\end{equation}
and
\begin{equation}
\vec{k}=\frac{1}{2}(\vec{p'}+\vec{p})
\end{equation}
instead of $\vec{p'}$ and $\vec{p}$ in the practical calculation.

In terms of the Feynman rules, the Dirac spinor
\begin{equation}
u_(\vec{q},s)=\sqrt{\frac{E+M}{2M}}\begin{pmatrix}
                                     1 \\
                                     \vec{\sigma}\cdot\vec{q} / (E+M)
                                   \end{pmatrix}\xi_s,
\end{equation}
\begin{equation}
\bar u_(\vec{q},s)=\xi^{\dagger}_s\sqrt{\frac{E+M}{2M}}\begin{pmatrix}
                                     1  &&
                                      -\vec{\sigma}\cdot\vec{q} / (E+M)
                                   \end{pmatrix}
\end{equation}
and the nonrealistic reduction
\begin{equation}
\frac{E}{m}\approx 1+\frac{\vec{p}^2}{2m^2},
\end{equation}
the scattering amplitude caused by the $\pi$-meson exchange can be written as
\begin{eqnarray}
iM_\pi=\!\!\!\!&&3g_{\pi NN}g_{\pi N^*N^*}\frac{1}{4 m_1 m_2} \frac{(\vec{\sigma_1}\cdot\vec{q})(\vec{\sigma_2}
\cdot\vec{q})}{\vec{q}^{2}+m^{2}_\pi}
+3g^{2}_{\pi N^*N^*}\frac{(m_1+m_2)^2}{8 m^{2}_1 m^{2}_2} \frac{(\vec{\sigma_1}\cdot\vec{q})(\vec{\sigma_2}
\cdot\vec{q})}{\vec{q}^{2}-M^{2}_\pi},
\end{eqnarray}
and the scattering amplitude induced by the $\sigma$-meson exchange can be expressed by
\begin{eqnarray}
iM_\sigma=\!\!\!\!&&-g_{\sigma NN^*}g_{\sigma N^*N^*}\frac{1}{\vec{q}^{2}+m^{2}_\sigma}[1-\frac{(m_1+m_2)^2}{4 m^{2}_1 m^{2}_2}\vec{k}^{2}+
\frac{(m_1+m_2)^2}{16 m^{2}_1 m^{2}_2}\vec{q}^{2}-\frac{(m_1+m_2)^2}{4 m^{2}_1 m^{2}_2}i\vec{S}\cdot(\vec{q}\times\vec{k})]\nonumber\\
&&-g^2_{\sigma NN^*(1440)}\frac{1}{\vec{q}^{2}+M^{2}_\sigma}[1-\frac{1}{2 m_1 m_2}\vec{k}^{2}+
\frac{1}{8 m_1 m_2}\vec{q}^{2}-\frac{1}{2 m_1 m_2}i\vec{S}\cdot(\vec{q}\times\vec{k})].
\end{eqnarray}
The scattering amplitudes arisen from the $\omega$- and $\rho$-meson exchanges can be derived in the same way. Their forms are quite similar to those of the $\pi$- and $\sigma$-meson exchanges. To reduce the length of the paper, they will not be presented here. It should be mentioned that because of the mass difference between $N$ and $N^*$, there would be a so-called "{\it effective mass}" $M_{\pi,\sigma,\rho,\omega}$ in the amplitude formula
\begin{equation}
M^2_{\pi}=(m_{N^*(1440)}-m_N)^2-m^2_{\pi},
\end{equation}
\begin{equation}
M^2_{\sigma,\rho,\omega}=m^2_{\sigma,\rho,\omega}-(m_{N^*(1440)}-m_N)^2.
\end{equation}
Moreover, in the derivation, the term with symmetric combination $(\vec{\sigma_1}+\vec{\sigma_2})$ which leads to a spin-orbital term $\vec{S}\cdot(\vec{q}\times\vec{k})$ is remained, while the term with asymmetric combination $(\vec{\sigma_1}-\vec{\sigma_2})$ is dropped. Substituting $M_{fi}$ into Eq.(\ref{eq:vq}), a meson-exchange potential in the momentum space, $V(\vec{q},\vec{k})$, can be obtained.

Further making Fourier-transformation, the $N$-$N^*$(1440) potential in the coordinate space, $V_M(r)$, can be derived. It should be pointed out that because the mass of the $\pi$-meson is relatively small, the Fourier transformation of the $\pi$-meson-exchange potential would be a complex function, which implies that the off-shell effect of the exchanged pion would induce other reaction channels. Since what we are interested in is the elastic scattering, as an approximation, the imaginary part of the potential can temporarily be dropped, which is consistent with the treatment in Ref.\cite{X.liu}. For instance,
\begin{equation}
\mathcal{F}\{\frac{(\vec{\sigma_1}\cdot\vec{q})(\vec{\sigma_2}\cdot\vec{q})}
{\vec{p}^2-M^2}\}\cong-\frac{M}{3}
[~-\vec{\sigma_1}\cdot\vec{\sigma_2}M^2 \frac{\cos(Mr)}{Mr} - S_{12} M^2 Z'(Mr)~]
\end{equation}
with
\begin{equation}
Z'(Mr)=\frac{\cos(Mr)}{Mr}-\frac{3}{Mr}\frac{\sin(Mr)}{Mr}-\frac{3}
{(Mr)^2}\frac{\cos(Mr)}{Mr}.
\end{equation}

Considering the finite size effect of $N$ ($N^*$), similar to the derivation of the Bonn potential, a form factor
\begin{equation}
F(q)=\frac{\Lambda^2-m^2}{\Lambda^2-q^2}=(\frac{\Lambda^2-m^2}{\tilde{\Lambda}^2+
\vec{q}^2})
\label{eq:cut1}
\end{equation}
with $\Lambda$ being the cutoff parameter and
\begin{equation}
\tilde{\Lambda}^2=\Lambda^2+M^2-m^2
\label{eq:cut2}
\end{equation}
is adopted to suppress the high momentum contribution in the relatively lower energy scattering. Then the ultraviolet divergence in the Fourier transformation can be eliminated, and the potential in the coordinate space at small $r$ is remained finite. Finally, the potential of the $N$-$N^*$(1440) interaction can be written as
\begin{equation}
V(r)=V_{\pi}+V_{\sigma}+V_{\rho}+V_{\omega}=V_{C}+V_{LS}+V_{T}, \label{eq:totv}
\end{equation}
where $V_{C}$, $V_{LS}$ and $V_{T}$ denote the total central, spin-orbital and tensor potentials,
\begin{eqnarray}
V_\pi(r)=&&\frac{g_{\pi NN}g_{\pi N^*N^*}}{4\pi}\frac{3}{4m_1m_2}
[~(\vec{\sigma_1}\cdot\vec{\sigma_2})\mathcal{F}_{3t1} + S_{12}\mathcal{F}_{3t2}~]\nonumber\\
&&+\frac{g^{2}_{\pi NN^*(1440)}}{4\pi}\frac{3(m_1+m_2)^2}{16m^{2}_1m^{2}_2}
[~(\vec{\sigma_1}\cdot\vec{\sigma_2})\mathcal{F}_{6u1} + S_{12}\mathcal{F}_{6u2}~],
\end{eqnarray}
\begin{eqnarray}
V_\sigma(r)=&&\frac{g_{\sigma NN}g_{\sigma N^*N^*}}{4\pi}[~-\mathcal{F}_{1t}+
\frac{m^{2}_1+m^{2}_2}{4m^{2}_1m^{2}_2}\mathcal{F}_{4t}
-\frac{m^{2}_1+m^{2}_2}{16m^{2}_1m^{2}_2}\mathcal{F}_{2t}
+\frac{m^{2}_1+m^{2}_2}{4m^{2}_1m^{2}_2}\vec{S}\cdot\vec{L}\mathcal{F}_{5t0}~]\nonumber\\
&&+\frac{g^{2}_{\sigma NN^*}}{4\pi}[~-\mathcal{F}_{1u}+
\frac{1}{2m_1m_2}\mathcal{F}_{4u}
-\frac{1}{8m_1m_2}\mathcal{F}_{2u}
+\frac{1}{2m_1m_2}\vec{S}\cdot\vec{L}\mathcal{F}_{5u0}~],
\end{eqnarray}
\begin{eqnarray}
V_\rho(r)=&&\frac{g_{\rho NN}g_{\rho N^*N^*}}{4\pi}[~-3\mathcal{F}_{1t}-
\frac{3m^{2}_1+3m^{2}_2+12m_1m_2}{4m^{2}_1m^{2}_2}\mathcal{F}_{4t}
+\frac{3m^{2}_1+3m^{2}_2}{16m^{2}_1m^{2}_2}\mathcal{F}_{2t}\nonumber\\
&&-\frac{3m^{2}_1+3m^{2}_2+12m_1m_2}{4m^{2}_1m^{2}_2}\vec{S}\cdot\vec{L}\mathcal{F}_{5t0}
+\frac{3\vec{\sigma_1}\cdot\vec{\sigma_2}}{2m_1m_2}(\mathcal{F}_{2t}-\mathcal{F}_{3t1})
-\frac{3}{2m_1m_2}S_{12}\mathcal{F}_{3t2}~]\nonumber\\
&&+\frac{f_{\rho NN^*}}{2m_1+2m_2}\frac{g_{\rho NN^*}}{4\pi}[~\frac{3}{m_1}\mathcal{F}_{2t}
-\frac{6(m_1+m_2)}{m_1m_2}\vec{S}\cdot\vec{L}\mathcal{F}_{5t0}+\frac{3\vec{\sigma_1}\cdot\vec{\sigma_2}}{m_2}
(\mathcal{F}_{2t}-\mathcal{F}_{3t1})-\frac{3S_{12}}{m_2}\mathcal{F}_{3t2}~]\nonumber\\
&&+\frac{f_{\rho N^*N^*}}{2m_1+2m_2}\frac{g_{\rho NN}}{4\pi}[~\frac{3}{m_2}\mathcal{F}_{2t}
-\frac{6(m_1+m_2)}{m_1m_2}\vec{S}\cdot\vec{L}\mathcal{F}_{5t0}+\frac{3\vec{\sigma_1}\cdot\vec{\sigma_2}}{m_1}
(\mathcal{F}_{2t}-\mathcal{F}_{3t1})-\frac{3S_{12}}{m_1}\mathcal{F}_{3t2}~]\nonumber\\
&&+\frac{f_{\rho N^*N^*}}{(2m_1+2m_2)^2}\frac{f_{\rho NN}}{4\pi}[~12\vec{\sigma_1}\cdot\vec{\sigma_2}
(\mathcal{F}_{2t}-\mathcal{F}_{3t1})-12S_{12}\mathcal{F}_{3t2}~]\nonumber\\
&&\frac{g^{2}_{\rho N^*N}}{4\pi}[~-3\mathcal{F}_{1u}-
\frac{3m^{2}_1+3m^{2}_2+12m_1m_2}{4m^{2}_1m^{2}_2}\mathcal{F}_{4u}
+\frac{3m^{2}_1+3m^{2}_2}{16m^{2}_1m^{2}_2}\mathcal{F}_{2u}\nonumber\\
&&-\frac{3m^{2}_1+3m^{2}_2+3m_1m_2}{2m^{2}_1m^{2}_2}\vec{S}\cdot\vec{L}\mathcal{F}_{5u0}
-\frac{3(m_1-m_2)^2}{4m^{2}_1m^{2}_2}\vec{\sigma_1}\cdot\vec{\sigma_2}\mathcal{F}_4u\nonumber\\
&&+\frac{3(m_1+m_2)^2}{16m^{2}_1m^{2}_2}\vec{\sigma_1}\cdot\vec{\sigma_2}
(\mathcal{F}_{2u}-\mathcal{F}_{3u1})-\frac{3(m_1+m_2)^2}{16m^{2}_1m^{2}_2}S_{12}\mathcal{F}_{3u2}~]\nonumber\\
&&\frac{g^{2}_{\rho N^*N}}{4\pi}\frac{(m_1-m_2)^2}{m^{2}_\rho}[~3\mathcal{F}_{1u}-
\frac{3}{2m_1m_2}\mathcal{F}_{4u}+\frac{3}{8m_1m_2}\mathcal{F}_{2u}
-\frac{3}{2m_1m_2}\vec{S}\cdot\vec{L}\mathcal{F}_{5u0}~]\nonumber\\
&&+\frac{f_{\rho NN^*}}{2m_1+2m_2}\frac{g_{\rho NN^*}}{4\pi}[~
-\frac{3(m_1-m_2)^2(m_1+m_2)}{m^{2}_1m^{2}_2}\mathcal{F}_{4u}
+\frac{3(m_1+m_2)^3}{4m^{2}_1m^{2}_2}\mathcal{F}_{2u}\nonumber\\
&&-\frac{6(m_1+m_2)(m^{2}_1+m^{2}_2)}{m^{2}_1m^{2}_2}\vec{S}\cdot\vec{L}\mathcal{F}_{5u0}
-\frac{3(m_1+m_2)(m_1-m_2)^{2}}{m^{2}_1m^{2}_2}\vec{\sigma_1}\cdot\vec{\sigma_2}\mathcal{F}_{4u}\nonumber\\
&&+\frac{3(m_1+m_2)^3}{4m^{2}_1m^{2}_2}\vec{\sigma_1}\cdot\vec{\sigma_2}(\mathcal{F}_{2u}-\mathcal{F}_{3u1})
-\frac{3(m_1+m_2)^3}{4m^{2}_1m^{2}_2}S_{12}\mathcal{F}_{3u2}~]\nonumber\\
&&+\frac{f^{2}_{\rho N^*N}}{4\pi(2m_1+2m_2)^2}[~-\frac{3(m_1-m_2)^{4}}{m^{2}_1m^{2}_2}\mathcal{F}_{4u}
+\frac{3(m_1+m_2)^2(m_1-m_2)^2}{4m^{2}_1m^{2}_2}\mathcal{F}_{2u}\nonumber\\
&&-\frac{6(m_1+m_2)^2(m_1-m_2)^2}{m^{2}_1m^{2}_2}\vec{S}\cdot\vec{L}\mathcal{F}_{5u0}
-\frac{3(m_1+m_2)^2(m_1-m_2)^{2}}{m^{2}_1m^{2}_2}\vec{\sigma_1}\cdot\vec{\sigma_2}\mathcal{F}_{4u}\nonumber\\
&&+\frac{3(m_1+m_2)^{4}}{4m^{2}_1m^{2}_2}\vec{\sigma_1}\cdot\vec{\sigma_2}(\mathcal{F}_{2u}-\mathcal{F}_{3u1})
-\frac{3(m_1+m_2)^{4}}{4m^{2}_1m^{2}_2}S_{12}\mathcal{F}_{3u2}~]
\end{eqnarray}
and
\begin{eqnarray}
V_\omega(r)=-\frac{1}{3}V_\rho(r)(&&\!\!\!\!m_\rho\rightarrow m_\omega,f_{\rho NN^*}\rightarrow f_{\omega NN^*},
f_{\rho N^*N^*}\rightarrow f_{\omega N^*N^*},f_{\rho NN}\rightarrow f_{\omega NN},\nonumber\\
&&g_{\rho NN^*}\rightarrow g_{\omega NN^*},
g_{\rho N^*N^*}\rightarrow g_{\omega N^*N^*},g_{\rho NN}\rightarrow g_{\omega NN})
\end{eqnarray}
represent the potentials induced by the $\pi$-, $\sigma$-, $\rho$- and $\omega$-meson exchanges, respectively. The explicit forms of $\mathcal{F}_{\alpha t}$, $\mathcal{F}_{\alpha u}$, $\mathcal{F}_{\alpha t \beta}$ and $\mathcal{F}_{\alpha u \beta}$ are shown in the Appendix.

As to the cutoff parameters $\Lambda_\alpha$ (i.e. $\Lambda$ in Eqs.(\ref{eq:cut1},\ref{eq:cut2})), it is reasonable to take the values used in the Bonn $N$-$N$ potential temporarily at the very beginning ($\Lambda_{\pi}=1.3GeV$, $\Lambda_{\sigma}=1.1GeV$, $\Lambda_{\rho}=1.3GeV$ and $\Lambda_{\omega}=1.5GeV$, which is called Set $A_{\Lambda}$)~\cite{R.machleidt1987}, because $N^*$(1440) is an excited state of $N$ and $NN^*$(1440) has the same quantum numbers as the deuteron. However, comparing with $N$, $N^*$(1440) has one $p$-wave excitation, we should check the rationality of this adoption, re-adjust $\Lambda_\alpha$s to some reasonable values (we will call it as Set $B_{\Lambda}$) and use them in the later calculation.

\section{Coupling Constants}

Before realistic calculation, 18 coupling constants should be fixed, among them $g_{\pi NN}$, $g_{\sigma NN}$, $g_{\rho NN}$, $g_{\omega NN}$, $f_{\rho NN}$ and $f_{\omega NN}$ for meson-$N$-$N$ couplings, $g_{\pi NN^*(1440)}$, $g_{\sigma NN^*(1440)}$, $g_{\rho NN^*(1440)}$, $g_{\omega NN^*(1440)}$, $f_{\rho NN^*(1440)}$ and $f_{\omega NN^*(1440)}$ for meson-$N$-$N^*$(1440) couplings and $g_{\pi N^*N^*}$, $g_{\sigma N^*N^*}$, $g_{\rho N^*N^*}$, $g_{\omega N^*N^*}$, $f_{\rho N^*N^*}$ and $f_{\omega N^*N^*}$ for meson-$N^*$(1440)-$N^*$(1440) couplings, respectively.

The first 6 coupling constants, $g_{\pi NN}$, $g_{\sigma NN}$, $g_{\rho NN}$, $g_{\omega NN}$, $f_{\rho NN}$ and $f_{\omega NN}$, can be taken from the Bonn potential~\cite{R.machleidt1987}, because the $\delta$- and $\eta$-meson exchanges do not contribute due to their strange constituents.

\begin{figure}[htbp]
  \begin{center}
 \includegraphics*[width=0.32\textwidth]{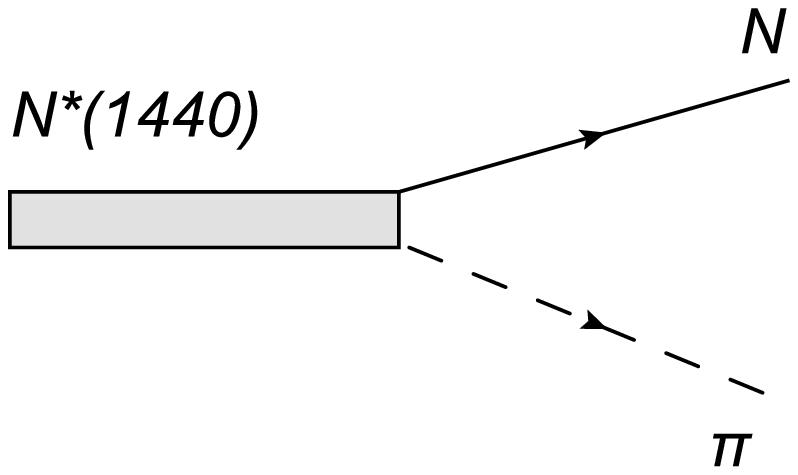}
    \includegraphics*[width=0.32\textwidth]{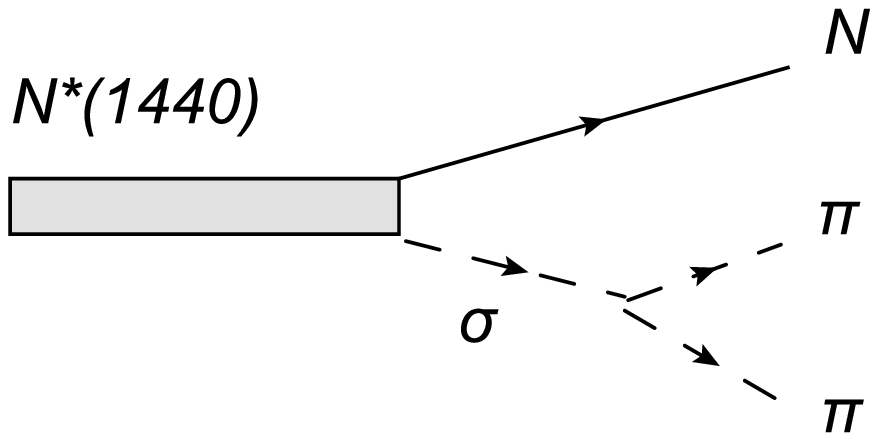}
    \caption{ Feynman diagrams for the $N^*(1440)\rightarrow N \pi$  (left)
    and $N^*(1440)\rightarrow N \sigma \rightarrow N \pi \pi$ (right).}
    \label{width15}
  \end{center}
\end{figure}

The coupling constants for the meson-$N$-$N^*$(1440) interaction can phenomenologically be extracted from the relevant data fitting. More specifically, $g_{\pi NN^*(1440)}$ can be obtained by fitting the partial decay width of $N^*$(1440) to $N\pi$, namely the total width $\Gamma_{N^*(1440)}$ of $200\sim450(\approx 300) MeV$ times the branching ratio $BR(N^*(1440)\to N\pi)$ of $0.55\sim0.75$~\cite{pdg}. The corresponding Feynman diagram is shown in the left diagram in Fig.\ref{width15}.
Based on this diagram, the coupling constant can be derived as
\begin{eqnarray}
\frac{g^{2}_{\pi NN^*(1440)}}{4\pi}=\frac{4 m^{3}_{N^*}\Gamma_{N^*\rightarrow N \pi}}{3[(m_{N^*}+m_N)^2-m^{2}_{\pi}]^{1/2}
[(m_{N^*}-m_N)^2-m^{2}_{\pi}]^{3/2}},\label{eq:npi}
\end{eqnarray}
with $\Gamma_{N^*\rightarrow N \pi}=\Gamma_{N^*(1440)}\times BR(N^*(1440)\to N\pi)$ being the partial decay width. Substituting the $\Gamma_{N^*\rightarrow N \pi}$ data into Eq.(\ref{eq:npi}), the value of $g_{\pi NN^*(1440)}$ can be extracted. They are tabulated in Table~\ref{tab:couplingA}.
\begin{table}[htbp]
\centering \caption{$g_{\alpha NN^*(1440)}$, $f_{\alpha NN^*(1440)}$ and $\Lambda_\alpha$ with $\alpha=\pi(138MeV)$, $\sigma(550MeV)$, $\rho(775MeV)$ and $\omega(782MeV)$ in Set $A_{\Lambda}$.}
 \label{tab:couplingA}
\begin{center}
\begin{tabular}{c c c c c c}
\hline\hline
\multirow{2}{*}{Coupling mode of $\rho$~~~~~} & \multirow{2}{*}{~~~Coupling constant~~~~~~}  &
\multicolumn{4}{c}{$\alpha$}  \\
\cline{3-6}
   &   & $~~~~~~\pi~~~~~~$ & $~~~~~~\sigma~~~~~~$  & $~~~~~~\rho~~~~~~$ & $~~~~~~\omega~~~~~~$ \\
\hline
\multirow{2}{*}{V+T} & $g^{2}_{\alpha NN^*(1440)}/{4\pi}$ & 2.9087  & 5.1200             &  0.3685 & 1.3890\\
\cline{2-6}
 & $f^{2}_{\alpha NN^*(1440)}/{4\pi}$ & --- & --- &  2.8288 & ---\\
\hline
\multirow{2}{*}{V} & $g^{2}_{\alpha NN^*(1440)}/{4\pi}$ & 2.9087    & 5.1200             &  1.1374 & 1.3890\\
\cline{2-6}
 & $f^{2}_{\alpha NN^*(1440)}/{4\pi}$ & --- & --- &  --- & ---\\
\hline
\multirow{2}{*}{T} & $g^{2}_{\alpha NN^*(1440)}/{4\pi}$ & 2.9087    & 5.1200             &  --- & 1.3890\\
\cline{2-6}
 & $f^{2}_{\alpha NN^*(1440)}/{4\pi}$ & --- & --- &  1.1374 & ---\\
\hline\hline
\multirow{2}{*}{V+T} & $g^{2}_{\alpha NN}/{4\pi}$ & 14.900  & 7.7823 &  0.9500 & 20.000\\
\cline{2-6}
 & $f^{2}_{\alpha NN}/{4\pi}$ & --- & --- &  35.350 & ---\\
\hline
 & $\Lambda_\alpha$ (GeV) & 1.3000 & 1.1000 &  1.3000 & 1.5000\\
\hline
\hline
\end{tabular}
\end{center}
\end{table}

In the same way, $g_{\sigma NN^*(1440)}$ can be extracted by fitting the branching ratio $BR(N^*(1440)\rightarrow N \sigma \rightarrow N \pi \pi)$ of $0.11\sim0.13$~\cite{pdg}. The corresponding Feynman diagram is shown in the right diagram in Fig.\ref{width15}. Based on this diagram, the coupling constant can be written as
\begin{eqnarray}
\frac{g^{2}_{\sigma NN^*(1440)}}{4\pi}=\frac{\Gamma_{N^*\rightarrow N \pi\pi}}{g^{2}_{\sigma \pi\pi}}
\frac{8\pi^4 m_{N^*}}{\mathcal{E}}
\end{eqnarray}
with factor ${\mathcal{E}}$ being
\begin{eqnarray}
\mathcal{E}=\int \frac{\Lambda^2-m^{2}_{\sigma}}{\Lambda^2-(p_3+p_4)^2}
\frac{(p_3\cdot p_4)^2(p\cdot p_1+ m_{N^*}m_N)}{((p_3+p_4)^2-m^{2}_{\sigma})^2+m^{2}_{\sigma}\Gamma^{2}_{\sigma}}
\frac{d^3\vec{p_{1}} d^3\vec{p_{3}} d^3\vec{p_{4}}}{8 p_{1}^{0} p_{3}^{0} p_{4}^{0}} \delta^4(p-p_1-p_3-p_4).
\end{eqnarray}

\begin{figure}[htbp]
  \begin{center}
  \rotatebox{0}{\includegraphics*[width=0.4\textwidth]{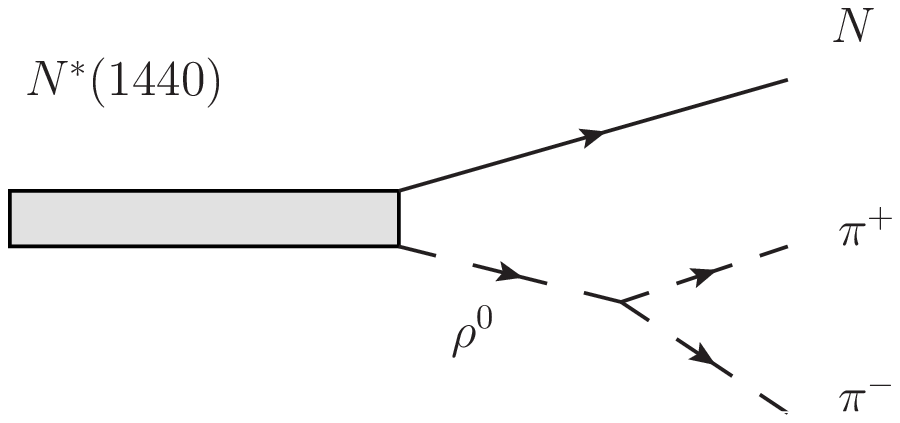}}
    \rotatebox{0}{\includegraphics*[width=0.4\textwidth]{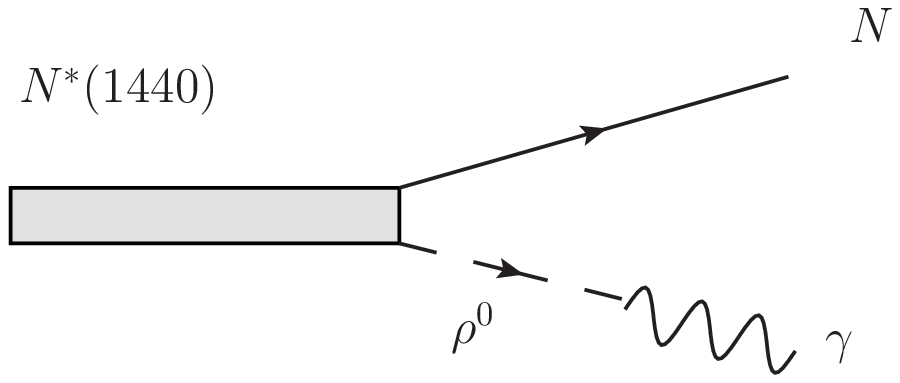}}
    \caption{ Feynman diagrams for the $N^*(1440)\rightarrow N \rho \rightarrow N \pi \pi$ (left)
    and $N^*(1440)\rightarrow p \gamma~(I_{\gamma}=1)$ (right).}
    \label{rho}
  \end{center}
\end{figure}

Extracting the value of the coupling constant for the $\rho$-meson is somewhat complicated. There are two types of coupling modes in the $\rho$-$N$-$N^*$ interaction, thus two coupling constants, the vector type coupling constant $g_{\rho NN^*(1440)}$ and the tensor type coupling constant $f_{\rho NN^*(1440)}$, should be fixed. To do so, we need two experimental values. Here, we choose the partial widths of the $N^*(1440)\to N\rho\to N\pi\pi$ and $N^*(1440)\to N\gamma~(I_{\gamma}=1)$ decays. The former width can be obtained by multiplying the total width $\Gamma_{N^*(1440)}$ of about 0.3 $GeV$ and the branching ratio $BR(N^*(1440)\to N\rho \to N\pi\pi)$ of $<8\%$~\cite{pdg}. In the practical calculation, taking $BR(N^*(1440)\to N\rho \to N\pi\pi)=2\%$ would be reasonable. The later width can be calculated by the following equation
\begin{eqnarray}
\Gamma_{N^* \rightarrow N \gamma~(I_{\gamma}=1)}= \frac{k^2}{4 \pi}\frac{m_N}{m_N^*}(A^{p \gamma}_{1/2}-A^{n \gamma}_{1/2})^2
\end{eqnarray}
with $A^{p \gamma}_{1/2} = -0.065 $ $GeV^{-1/2}$ and $A^{n
\gamma}_{1/2} = 0.04 $ $GeV^{-1/2}$ from the PDG~\cite{pdg},
respectively. The corresponding Feynman diagrams are shown in
Fig.\ref{rho}, where the vector-meson-dominant mechanism is employed
in the right diagram.
Based on this figure and Feynman rules, we can write the partial decay widths for these processes as
\begin{eqnarray}
\Gamma_{N^* \rightarrow N \pi^+ \pi^-}= && \int \delta^4(P-P_1-P_3-P_4)\frac{d^3p_1}{2P^0_{1}}\frac{d^3p_3}{2P^0_{3}}
\frac{d^3p_4}{2P^0_{4}} \frac{m_N}{(2\pi)^5}(\frac{\Lambda^2-m^2_\rho}{\Lambda^2-P^2_{2}})^2 \frac{1}{(P^2_{2}-m^2_\rho)^2}\nonumber\\
&&g^2_{\rho \pi\pi}[(g_{\rho NN^*} + f_{\rho NN^*})^2 c_1 + (2\frac{f_{\rho NN^*}}{m_{N^*}\!+\!m_N}P\cdot(P_4-P_3))^2 c_2]\label{eq:npipi}
\end{eqnarray}
and
\begin{eqnarray}
\Gamma_{N^* \rightarrow N \gamma~(I_{\gamma}=1)}=&& \frac{g^2_{\rho \gamma}}{4\pi}\frac{(m^2_{N^*} - m^2_N)(m_{N^*}- m_N)^2}{2 m^3_{N^*}}
(\frac{\Lambda^2-m^2_\rho}{\Lambda^2})^2 \frac{1}{m^4_{\rho}}(g_{\rho NN^*} + f_{\rho NN^*})^2\label{eq:ngamma}
\end{eqnarray}
where the factors $c_i$ can be written as
\begin{eqnarray}
c_1=\frac{2[P\cdot(P_4-P_3)][P_1\cdot(P_4-P_3)]-(P\cdot P_1)\cdot(P_4-P_3)^2 }{m_{N^*}m_N} + (P_4-P_3)^2
\end{eqnarray}
and
\begin{eqnarray}
c_2=\frac{(P\cdot P_1)}{m_{N^*}m_N} + 1,
\end{eqnarray}
respectively, and the $\rho \gamma$ coupling constants $g_{\rho \gamma}$ is given by ~\cite{photopro}
\begin{eqnarray}
g^2_{\rho \gamma}=\frac{e^2/4\pi\cdot m^4_{\rho}}{f^2_{\rho}/4\pi}
\end{eqnarray}
with $f^2_{\rho}/4\pi = 2.7$. Solving coupled Eqs.(\ref{eq:npipi}) and (\ref{eq:ngamma}), we can obtain the values of $g_{\rho NN^*(1440)}$ and $f_{\rho NN^*(1440)}$. We tabulate them in Table~\ref{tab:couplingA} as well.

To get more information about the $\rho$-$N$-$N^*$ coupling, we also arrange for the cases in which only one $\rho$-$N$-$N^*$ coupling mode, either the vector or the tensor, exists. In these cases, one value of the partial decay width would be enough to fix either $g_{\rho NN^*(1440)}$ or $f_{\rho NN^*(1440)}$. Then, the corresponding partial width formulas become
\begin{eqnarray}
\Gamma_{N^* \rightarrow N \gamma~(I_{\gamma}=1)}=&& g^2_{\rho NN^*} \frac{g^2_{\rho \gamma}}{4\pi}\frac{(m^2_{N^*} - m^2_N)(m_{N^*}- m_N)^2}{2 m^3_{N^*}}
(\frac{\Lambda^2-m^2_\rho}{\Lambda^2})^2 \frac{1}{m^4_{\rho}},\nonumber\\
\end{eqnarray}
and
\begin{eqnarray}
\Gamma_{N^* \rightarrow N \gamma~(I_{\gamma}=1)}=&& (\frac{f_{\rho NN^*}}{2m_{N^*}\!+\!2m_N})^2 \frac{g^2_{\rho \gamma}}{4\pi}\frac{(m^2_{N^*} - m^2_N)(m_{N^*}- m_N)^2}{2 m^3_{N^*}}(\frac{\Lambda^2-m^2_\rho}{\Lambda^2})^2 \frac{1}{m^4_{\rho}}.\nonumber\\
\end{eqnarray}
Substituting the measured partial decay width into the above equations, we obtain the value of $g_{\rho NN^*(1440)}$ or $f_{\rho NN^*(1440)}$. We also list them in Table~\ref{tab:couplingA}.

\begin{figure}[ht]
  \begin{center}
    \rotatebox{0}{\includegraphics*[width=0.5\textwidth]{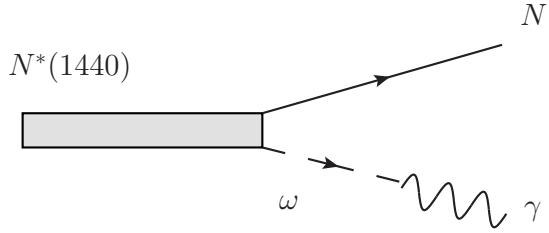}}
    \caption{Feynman diagrams for the $N^*(1440)\rightarrow p \gamma~(I_{\gamma}=0)$.\label{omega}}
  \end{center}
\end{figure}

Finally, we extract the coupling constant of the $\omega$-$N$-$N^*$(1440) interaction. As well known, the $\omega$-$N$-$N$ interaction has a vector type coupling only. Due to lack of $\Gamma_{N^*\rightarrow N \omega}$ data at the present moment, roughly taking a single vector coupling mode for the $\omega$-$N$-$N^*$(1440) interaction would make sense. Thus, we can use the partial decay width of the $N^*(1440)\to N\gamma~(I_{\gamma}=0)$ process with $A^{p \gamma}_{1/2} = -0.065 $ $GeV^{-1/2}$ and $A^{n \gamma}_{1/2} = 0.04 $  $GeV^{-1/2}$~\cite{pdg} to determine $g_{\omega NN^*(1440)}$. The corresponding Feynman diagram is shown in Fig.\ref{omega}.
According to this diagram, we can calculate the partial decay width by using
\begin{eqnarray}
\Gamma_{N^* \rightarrow N \gamma~(I_{\gamma}=0)}=&& \frac{k^2}{4 \pi}\frac{m_N}{m_N^*}(A^{p \gamma}_{1/2}+ A^{n \gamma}_{1/2})^2.
\end{eqnarray}
Meanwhile, we can write the partial decay width formula as
\begin{eqnarray}
\Gamma_{N^* \rightarrow N \gamma~(I_{\gamma}=0)}
=&& g^2_{\omega NN^*} \frac{g^2_{\omega \gamma}}{4\pi}\frac{(m^2_{N^*} - m^2_N)(m_{N^*}- m_N)^2}{2 m^3_{N^*}}
(\frac{\Lambda^2-m^2_\rho}{\Lambda^2})^2 \frac{1}{m^4_{\omega}}
\end{eqnarray}
with
\begin{eqnarray}
g^2_{\omega \gamma}=\frac{e^2/4\pi\cdot m^4_{\omega}}{f^2_{\omega}/4\pi}.
\end{eqnarray}
In the last equation, we roughly take $f^2_{\omega}/4\pi = 9f^2_{\rho}/4\pi= 24.3$ via SU(3) relation. By fitting measured partial decay width, we can extract the value of $g_{\omega NN^*(1440)}$ approximately, and also tabulate it in Table~\ref{tab:couplingA}.

However, we do not have any experimental data at all to determine the coupling constants of the meson-$N^*$-$N^*$ interaction at the present stage. Therefore, we have to surmise a way to set the values of $g_{\alpha N^*N^*}, (\alpha=\pi,~\sigma,~\rho,~\omega)$ and $f_{\beta N^*N^*}, (\beta=\rho,~\omega)$. Here, we consider two cases: Case 1, $g_{\alpha N^*N^*}=g_{\alpha NN}, (\alpha=\pi,~\sigma,~\rho,~\omega)$ and $f_{\beta N^*N^*}=f_{\beta NN}, (\beta=\rho,~\omega)$; and Case 2, $g_{\alpha N^*N^*}=g_{\alpha NN^*}, (\alpha=\pi,~\sigma,~\rho,~\omega)$ and $f_{\beta N^*N^*}=f_{\beta NN^*}, (\beta=\rho,~\omega)$. In each case, we have three coupling modes for the $\rho-N-N^*$ interaction, namely the vector+tensor coupling mode(VT), the vector coupling mode only (V) and the tensor coupling mode only (T).

Based on the uncertainties of the data, both the total width of $N^*$(1440) and the measured branching ratios for the above mentioned processes, we further tabulate corresponding partial widths and, consequently, the extended ranges of the extracted coupling constants in Table~{\ref{tab:gvar}}.
\begin{table}[htbp]
\centering \caption{Uncertainties of partial decay widths in concerned decay processes and corresponding variational ranges of $g_{\alpha NN^*(1440)}$ and $f_{\alpha NN^*(1440)}$ for $\pi$, $\sigma$, $\rho$ and $\omega$ with the values of $\Lambda_{\alpha}$ in Set $A_{\Lambda}$.}
 \label{tab:gvar}
\begin{center}
\begin{tabular}{c c c c c }
\hline\hline
$~~~~~~~~~\alpha~~~~~~~~~$ &  $~~~~~~~~~\pi~~~~~~~~~$ & $~~~~~~~~~\sigma~~~~~~~~~$  & $~~~~~~~~~\rho~~~~~~~~~$ & $~~~~~~~~~\omega~~~~~~~~~$ \\
\hline
\multirow{2}{*}{branching ratio~~~~} & $~~~~\Gamma_{\pi}/\Gamma~~~~$ & $~~~~\Gamma_{\sigma}/\Gamma~~~~$  & $~~~~\Gamma_{N^*\to N\gamma~(I_{\gamma}=1)}~~~~$  &  $\Gamma_{N^*\to N\gamma~(I_{\gamma}=0)}$\\
\cline{2-5}
 & $0.55\sim 0.75$ & $0.11\sim 0.13$ & $0.074\sim 0.13$ &  $0.0011\sim 0.014$\\
\hline\hline
$g^{2}_{\alpha NN^*(1440)}/{4\pi}$ & $2.461\sim 3.356~~$ & $~~4.693\sim 5.547$    & $0.8543\sim 1.461$ &  $0.2689\sim 3.380$\\
\hline
$f^{2}_{\alpha NN^*(1440)}/{4\pi}$ & --- & ---    & $0.8543\sim 1.461$ &  ---\\
\hline\hline
\end{tabular}
\end{center}
\end{table}

\begin{figure}[htbp]
  \begin{center}
  \rotatebox{0}{\includegraphics*[width=0.45\textwidth]{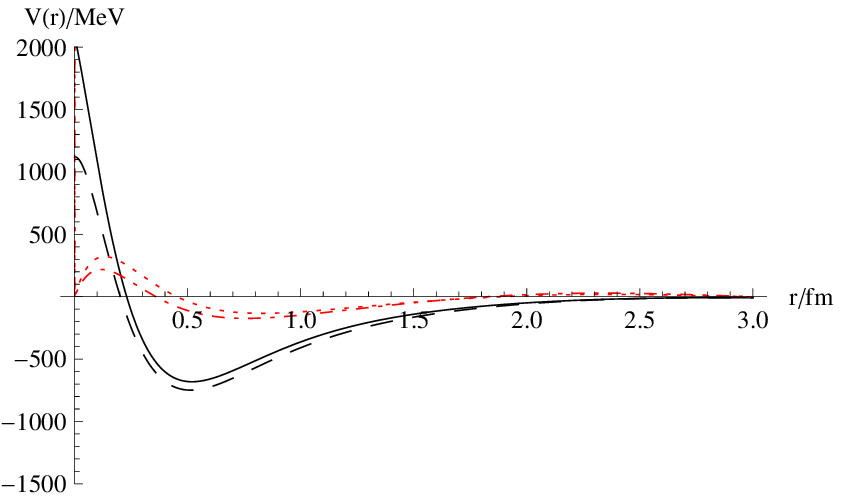}}
   \rotatebox{0}{\includegraphics*[width=0.45\textwidth]{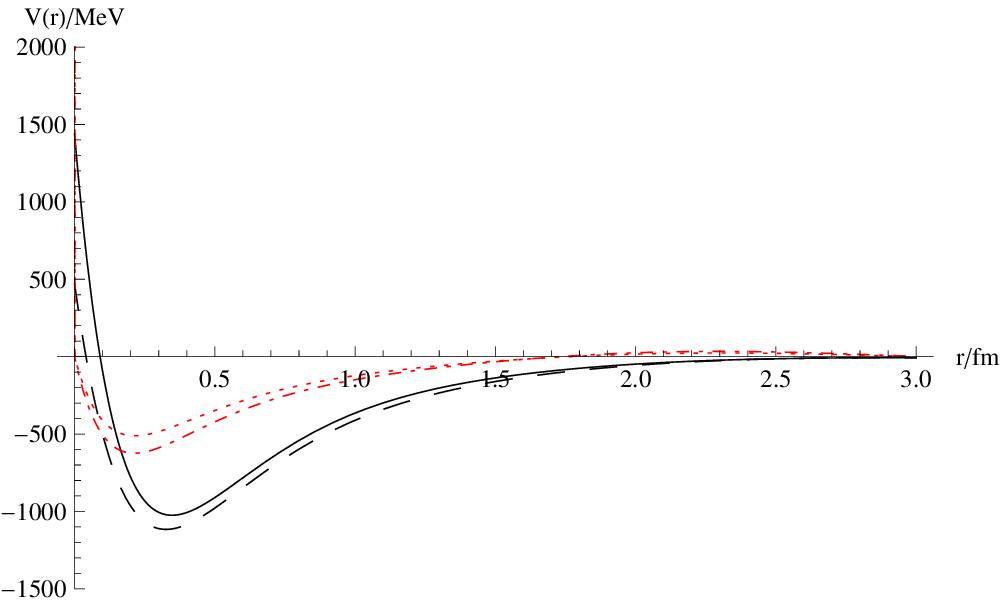}}
    \caption{$N$-$N^*$(1440) potentials in the cases 1V (left) and 2V (right). The solid and dashed curves denote the upper and lower bounds of the central potential, while the dotted and dash-dotted curves represent the upper and lower bounds of the tensor potential.}
    \label{central potential v}
  \end{center}
\end{figure}

\begin{figure}[htbp]
  \begin{center}
   \rotatebox{0}{\includegraphics*[width=0.45\textwidth]{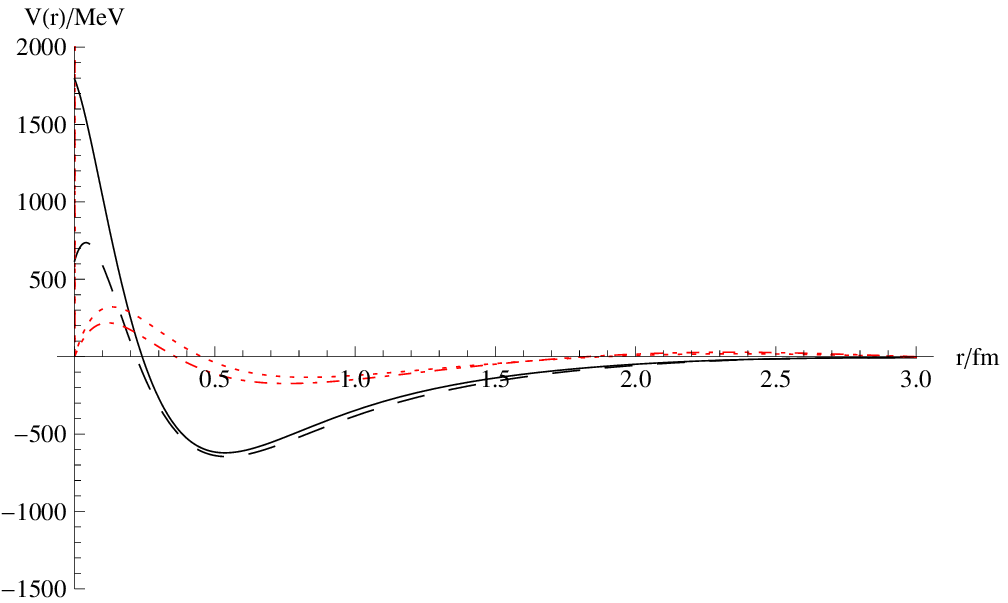}}
   \rotatebox{0}{\includegraphics*[width=0.45\textwidth]{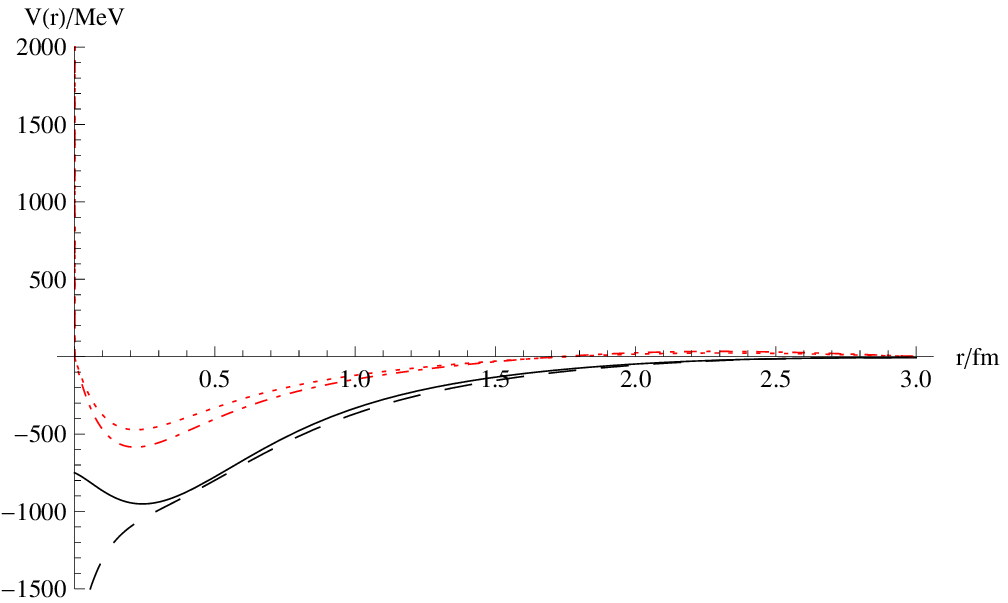}}
    \caption{$N$-$N^*$(1440) potentials in the cases 1T (left) and 2T (right). The solid and dashed curves denote the upper and lower bounds of the central potential, while the dotted and dash-dotted curves represent the upper and lower bounds of the tensor potential.}
    \label{central potential t}
  \end{center}
\end{figure}

\begin{figure}[htbp]
  \begin{center}
   \rotatebox{0}{\includegraphics*[width=0.45\textwidth]{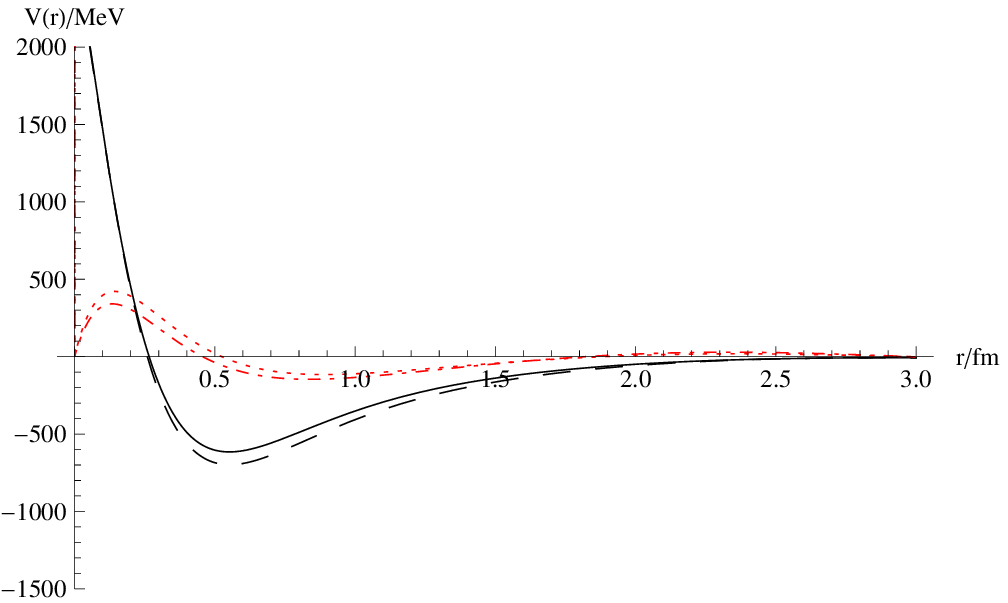}}
   \rotatebox{0}{\includegraphics*[width=0.45\textwidth]{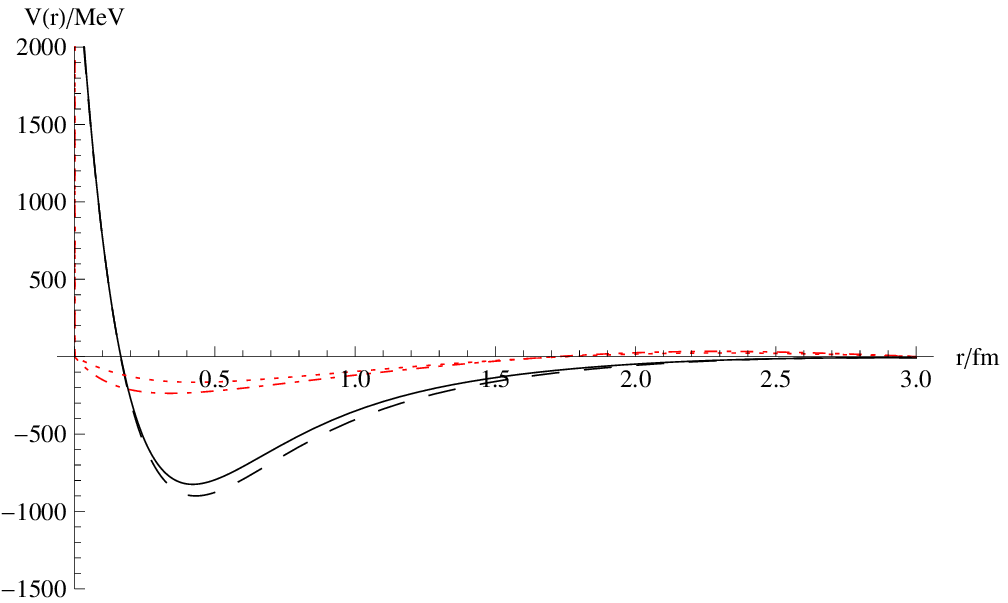}}
    \caption{$N$-$N^*$(1440) potentials in the cases 1VT (left) and 2VT (right).  The solid and dashed curves denote the upper and lower bounds of the central potential, while the dotted and dash-dotted curves represent the upper and lower bounds of the tensor potential. }
    \label{central potential vt}
  \end{center}
\end{figure}

To get a comprehensive impression, we plot the upper and lower bounds of the potentials, caused by the uncertainty of the partial decay width data, in the cases 1V and 2V in Fig.\ref{central potential v}, in the cases 1T and 2T in Fig.\ref{central potential t}, and in the cases 1VT and 2VT in Fig.\ref{central potential vt}, respectively. The solid and dashed curves denote the upper and lower bounds of the central potential, while the dotted and dash-dotted curves represent the upper and lower bounds of the tensor potential. From these curves, one sees that for the central potential, although a strong repulsion appears in the small $r$ region, an even stronger attractive effect in the medium and large $r$ regions might bring $N$ and $N^*$(1440) together to form a bound state. The tensor potentials in the case 2, no matter with which $\rho$ coupling mode, are stronger than those in the case 1, thus the $D$-state might couple to the $S$-state strongly, which is also beneficial for forming a bound state like in the deuteron case. Moreover, the strengths of the potentials in the cases with the T and VT coupling modes are closer but differ with that in the case with the V coupling mode, which implies that the tensor coupling mode in the $\rho$-$N$-$N^*$(1440) interaction plays a more important role than the vector mode does. Thus, the binding properties of the $N$-$N^*$(1440) system in the T and VT cases would be closer and would deviate from those in the V case.

\begin{figure}[ht]
  \begin{center}
  \rotatebox{0}{\includegraphics*[width=0.7\textwidth]{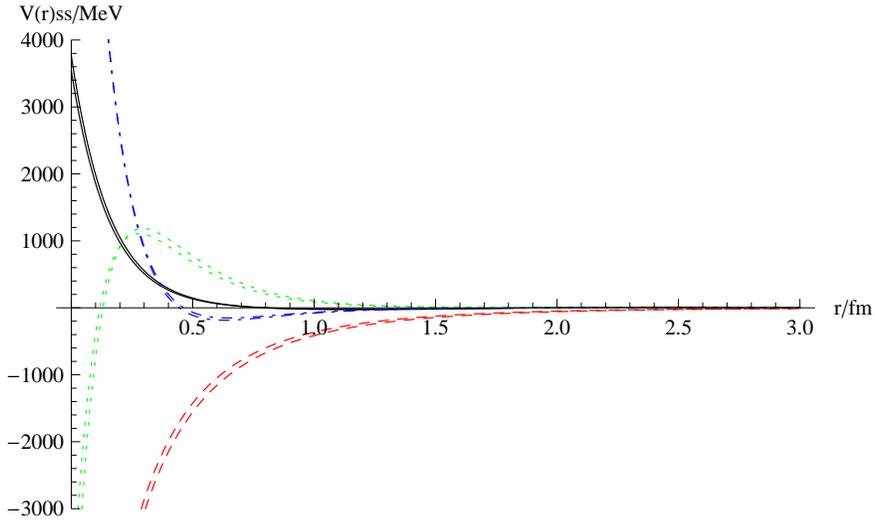}}
    \caption{ The upper and lower bounds of the central potentials contributed by various mesons for the case 1V. The solid, dashed, dash-dotted and dotted curves denote the contributions from the $\pi$-, $\sigma$-, $\rho$- and $\omega$-mesons, respectively.}
    \label{meson-potential}
  \end{center}
\end{figure}

We also plot the upper and lower bounds of the central potentials contributed by various mesons in the case 1V in Fig.~\ref{meson-potential}. The solid, dashed, dash-dotted and dotted curves denote the contributions from the $\pi$-, $\sigma$-, $\rho$- and $\omega$-mesons, respectively. Clearly, the $\sigma$-meson provides a major attractive force which plays a key role in binding, the $\rho$-meson gives a weak attraction in the medium $r$ region and a strong repulsion in the small $r$ region, the contribution from $\pi$-meson shows a very weak attractive feature in the long distance and a strong repulsive character in the short and medium distance, and the $\omega$-meson mainly provides a repulsion. The corresponding characters in the T and VT cases are similar to these, we would not show them again.

\section{Result and discussion}

Having the $N$-$N^*$(1440) potential $V(\vec{r})$ in Eq.(\ref{eq:totv}), we are able to study the binding property of the $N$-$N^*$(1440) system by solving the Schr\"{o}dinger Equation
\begin{equation}
(-\frac{\hbar^2}{2\mu}\nabla^{2}+V(\vec{r})-E)\Psi(\vec{r})=0, \label{eq:schrod}
\end{equation}
where $\Psi(\vec{r})$ is the total wave function of the system. The quantum numbers of the system we are going to study are the total spin $S=1$ and the orbital angular momenta $L=0$ and $L=2$. Thus the wave function $\Psi(\vec{r})$ should have a form of
\begin{equation}
\Psi(\vec{r})=\psi_S(\vec{r})+\psi_D(\vec{r}),
\end{equation}
where $\psi_S(\vec{r})$ and $\psi_D(\vec{r})$ are the $S$-wave and $D$-wave functions, respectively. In the matrix method, we use Laguerre polynomials as a set of orthogonal basis
\begin{equation}
\chi_{nl}(r)=\sqrt{\frac{(2\lambda)^(2l+3) n!}{\Gamma(2l+3+n)}}r^l e^{-\lambda r}L^{2l+2}_n (2\lambda r), n=1,2,3...
\end{equation}
with a normalization condition of
\begin{equation}
\int^\infty _0 \chi_{im}(r) \chi_{in}(r) r^2 dr=\delta_{ij}\delta_{mn},
\end{equation}
and expand the total wave function as
\begin{equation}
\Psi(\vec{r})=\sum^{n-1}_{i=0}a_i \chi_{i 0}(r)\phi_S + \sum^{n-1}_{p=0}b_p \chi_{p 2}(r)\phi_D, \label{eq:tottrail}
\end{equation}
where $\phi_S$ and $\phi_D$ are the angular part of the orbital wave function and spin wave function for the $S$- and $D$-states, respectively, and $a_i$ and $b_i$ are corresponding expansion coefficients.

In the practical calculation, Eq.(\ref{eq:schrod}) is re-written, by detaching the terms related to the kinetic-energy-operator ($\nabla^{2}$) from $V(\vec{r})$, as

\begin{equation}
(-\frac{\hbar^2}{2\mu}\nabla^{2}-\frac{\hbar^2}{2\mu}[\nabla^2
\alpha(r)+\alpha(r)\nabla^2]+\widetilde{V}(\vec{r})-E)\Psi(\vec{r})=0
\end{equation}
with
\begin{equation}
\nabla^2=\frac{1}{r}\frac{d^2}{dr^2}r-\frac{\overrightarrow{L}^2}{r^2},
\end{equation}
\begin{eqnarray}
\alpha(r)=&&(-2\mu)\{[~\frac{g_{\sigma NN}g_{\sigma N^*N^*}}{4\pi}
\frac{m^{2}_1+m^{2}_2}{4m^{2}_1m^{2}_2}\mathcal{F}_{4t2}
+\frac{g^{2}_{\sigma NN^*}}{4\pi}\frac{1}{2m_1m_2}\mathcal{F}_{4u2}~]\nonumber\\
&&+[~-\frac{g_{\rho NN}g_{\rho N^*N^*}}{4\pi}\frac{3m^{2}_1+3m^{2}_2+12m_1m_2}{4m^{2}_1m^{2}_2}\mathcal{F}_{4t2}\nonumber\\
&&+\frac{g^{2}_{\rho N^*N}}{4\pi}(-\frac{3m^{2}_1+3m^{2}_2+12m_1m_2}{4m^{2}_1m^{2}_2}
-\frac{3(m_1-m_2)^2}{4m^{2}_1m^{2}_2}\vec{\sigma_1}\cdot\vec{\sigma_2})\mathcal{F}_{4u2}\nonumber\\
&&-\frac{g^{2}_{\rho N^*N}}{4\pi}\frac{(m_1-m_2)^2}{m^{2}_\rho}\frac{3}{2m_1m_2}\mathcal{F}_{4u2}\nonumber\\
&&-\frac{f_{\rho NN^*}}{2m_1+2m_2}\frac{g_{\rho NN^*}}{4\pi}\frac{3(m_1-m_2)^2(m_1+m_2)}{m^{2}_1m^{2}_2}
(1+\vec{\sigma_1}\cdot\vec{\sigma_2})\mathcal{F}_{4u2}\nonumber\\
&&-\frac{f^{2}_{\rho N^*N}}{4\pi(2m_1+2m_2)^2}(\frac{3(m_1-m_2)^{4}}{m^{2}_1m^{2}_2}
+\frac{3(m_1+m_2)^2(m_1-m_2)^{2}}{m^{2}_1m^{2}_2}\vec{\sigma_1}\cdot\vec{\sigma_2})\mathcal{F}_{4u2}~]\nonumber\\
&&-\frac{1}{3}[~-\frac{g_{\omega NN}g_{\omega N^*N^*}}{4\pi}\frac{3m^{2}_1+3m^{2}_2+12m_1m_2}{4m^{2}_1m^{2}_2}\mathcal{F}_{4t2}\nonumber\\
&&+\frac{g^{2}_{\omega N^*N}}{4\pi}(-\frac{3m^{2}_1+3m^{2}_2+12m_1m_2}{4m^{2}_1m^{2}_2}
-\frac{3(m_1-m_2)^2}{4m^{2}_1m^{2}_2}\vec{\sigma_1}\cdot\vec{\sigma_2})\mathcal{F}_{4u2}\nonumber\\
&&-\frac{g^{2}_{\omega N^*N}}{4\pi}\frac{(m_1-m_2)^2}{m^{2}_\omega}\frac{3}{2m_1m_2}\mathcal{F}_{4u2}\nonumber\\
&&-\frac{f_{\omega NN^*}}{2m_1+2m_2}\frac{g_{\omega NN^*}}{4\pi}\frac{3(m_1-m_2)^2(m_1+m_2)}{m^{2}_1m^{2}_2}
(1+\vec{\sigma_1}\cdot\vec{\sigma_2})\mathcal{F}_{4u2}\nonumber\\
&&-\frac{f^{2}_{\omega N^*N}}{4\pi(2m_1+2m_2)^2}(\frac{3(m_1-m_2)^{4}}{m^{2}_1m^{2}_2}
+\frac{3(m_1+m_2)^2(m_1-m_2)^{2}}{m^{2}_1m^{2}_2}\vec{\sigma_1}
\cdot\vec{\sigma_2})\mathcal{F}_{4u2}~]\},
\end{eqnarray}
and the effective potential
\begin{equation}
\widetilde{V}(\vec{r})=V(\vec{r})-(terms~ related~ to~ \nabla^2).
\end{equation}

Then, with the wave function in Eq.(\ref{eq:tottrail}), the Hamiltonian matrix can be
expressed as
\begin{equation}
\begin{pmatrix}
H^{SS} & H^{SD}\\
H^{DS} & H^{DD}
\end{pmatrix}
\end{equation}
with
\begin{eqnarray}
H^{SS}=&&\langle \phi_S |\int^\infty _0 \sum^{n-1}_{i,j} a_i \chi_{i 0}(r) \{-\frac{\hbar^2}{2\mu}[1+\alpha(r)]\nabla^2 a_j \chi_{j 0}(r)
-\frac{\hbar^2}{2\mu}\nabla^2 [\alpha(r)a_j \chi_{j 0}(r)]\nonumber\\
&&+ V_{SS} (r) a_j \chi_{j 0}(r)\} r^2 dr | \phi_S \rangle,
\end{eqnarray}
\begin{eqnarray}
H^{SD}=\langle \phi_S |\int^\infty _0 \sum^{n-1}_{i,p} a_i \chi_{i
0}(r) V_{SD} (r) b_p \chi_{p 2}(r) r^2 dr | \phi_D \rangle,
\end{eqnarray}
\begin{eqnarray}
H^{DS}=\langle \phi_D |\int^\infty _0 \sum^{n-1}_{p,i} b_p \chi_{p
2}(r) V_{DS} (r)a_i \chi_{i 0}(r) r^2 dr | \phi_S \rangle,
\end{eqnarray}
and
\begin{eqnarray}
H^{DD}=&&\langle \phi_D |\int^\infty _0 \sum^{n-1}_{p,q} b_p \chi_{p
2}(r) \{-\frac{\hbar^2}{2\mu}[1+\alpha(r)]\nabla^2 b_q \chi_{q 2}(r)
-\frac{\hbar^2}{2\mu}\nabla^2 [\alpha(r)b_q \chi_{q 2}(r)]\nonumber\\
&&+ V_{DD} (r) b_q \chi_{q 2}(r)\} r^2 dr  | \phi_D \rangle.
\end{eqnarray}
In the above equations, $V_{SS}(r)$, $V_{SD}(r)$, $V_{DS}(r)$ and $V_{DD}(r)$ are  taken to be $\widetilde{V}_c(r)$, $\widetilde{V}_T(r)$, $\widetilde{V}_T(r)$ and $\widetilde{V}_c(r)+\widetilde{V}_T(r)$, respectively. Diagonalizing this matrix, the binding energy and corresponding wave function of the $NN^*$(1440) state can be obtained.

The resultant binding energies for the $S$-states with the coupling constants shown in Table~\ref{tab:couplingA} (with Set $A_{\Lambda}$) are tabulated in Table~\ref{tab:cen bound states A set}.
\begin{table}[htbp]
\caption{Binding energies of the bound $S$-states of the $NN^*$(1440) system (in MeV) with Set $A_{\Lambda}$.}
\label{tab:cen bound states A set}
\begin{center}
\begin{tabular}{c c c c}
\hline
\hline
\multirow{2}{*}{Coupling mode of $\rho$~~~~~} & \multirow{2}{*}{~~~~~~~~}  &
\multicolumn{2}{c}{Binding energy}  \\
\cline{2-4}
   &   &~~~~~~~~ case 1~~~~~~~~ &~~~~~~~~ case 2~~~~~~~~  \\
\hline
\multirow{2}{*}{V} & $E_1$ &   ~-373.7~     & ~-583.3~\\
\cline{2-4}
                    & $E_2$ & ~-23.2~     & ~-68.0~\\
\hline
\multirow{2}{*}{T} & $E_1$ &   ~-316.5~     & ~-422.5~\\
\cline{2-4}
                   & $E_2$ & ~-11.0~     & ~-29.0~\\
\hline
\multirow{2}{*}{V+T} & $E_1$ &  ~-324.6~    & ~-419.2~\\
\cline{2-4}
                     & $E_2$ & ~-13.0~     & ~-29.5~\\
\hline\hline
\end{tabular}
\end{center}
\end{table}
From this table, it is seen that with such a potential, no matter in which case and with which coupling mode, there might exist two bound $S$-states. One of them is a deeply bound state whose energy is close to the energy of the deuteron, and the other is a weakly bound state.

Contributions from various mesons to the binding energies of the obtained bound states in various cases with Set $A_{\Lambda}$ are shown in Table~\ref{tab:mcontri}. For simplicity,  relatively smaller contributions from the $D$-state are disregarded.
\begin{table}[htbp]
\caption{Contributions from various mesons and kinetic energy to the binding energies of the obtained bound states in various cases (in MeV)  with Set $A_{\Lambda}$.}
\label{tab:mcontri}
\begin{center}
\begin{tabular}{c c c c c c c c}
\hline
\hline
\multirow{2}{*}{~~~case~~~~~} & \multirow{2}{*}{~~~~~~~~}  &
\multicolumn{5}{c}{Binding energy}  \\
\cline{2-8}
   &   &~~~~~~~total~~~~~~~& kinetic energy &~~~$\pi$~~~&~~~~~$\sigma$~~~  &~~~$\rho$~~~   &~~~$\omega$~~~  \\
\hline
\multirow{2}{*}{case 1V} & $E_1$ &  ~-339.8~ & ~~~~146.7~~~~ &~~~~53.53~~~~ & ~~~~-821.4~~~~ & ~~~-31.17~~~& ~~~312.5~~~\\
\cline{2-8}
                         & $E_2$ &  ~-18.74~  & ~~~~101.4~~~~ & ~~~~13.00~~~~ &   ~-198.9~    & ~4.975~&    ~60.83~\\
\hline
\multirow{2}{*}{case 2V} & $E_1$ &  ~-477.3~ & ~~~~239.6~~~~ &~~~~83.75~~~~ & ~~~~-892.2~~~~ & ~~~-55.46~~~& ~~~146.9~~~\\
\cline{2-8}
                         & $E_2$ &  ~-48.50~  & ~~~~186.1~~~~ &~~~~14.34~~~~ &   ~-273.1~    & ~-7.307~&    ~31.47~\\
\hline
\multirow{2}{*}{case 1T} & $E_1$ &  ~-284.5~  & ~~~~132.1~~~~ &~~~~36.38~~~~ & ~~~~-707.3~~~~ & ~~~-6.001~~~& ~~~260.3~~~\\
\cline{2-8}
                         & $E_2$ &  ~-9.076~  & ~~~~56.46~~~~ & ~~~~6.346~~~~ &   ~-104.8~    & ~3.551~&    ~29.39~\\
\hline
\multirow{2}{*}{case 2T} & $E_1$ &  ~-363.2~ & ~~~~210.3~~~~ &~~~~56.74~~~~ & ~~~~-771.2~~~~ & ~~~-19.06~~~& ~~~121.8~~~\\
\cline{2-8}
                         & $E_2$ &  ~-22.74~  & ~~~~112.4~~~~ & ~~~~~8.051~~~~~~ &   ~-165.5~    & ~4.256~&    ~18.04~\\
\hline
\multirow{2}{*}{case 1VT} & $E_1$ &  ~-297.8~ & ~~~~131.9~~~~ &~~~~36.44~~~~ & ~~~~-715.7~~~~ & ~~~-15.71~~~& ~~~265.3~~~\\
\cline{2-8}
                         & $E_2$ &  ~-11.11~  & ~~~~65.35~~~~ & ~~~~7.407~~~~ &   ~-123.6~    & ~3.877~&    ~35.89~\\
\hline
\multirow{2}{*}{case 2VT} & $E_1$ &  ~-379.9~ & ~~~~182.7~~~~ &~~~45.34~~~~ & ~~~~-731.2~~~~ & ~~~7.715~~~& ~~~115.5~~~\\
\cline{2-8}
                         & $E_2$ &  ~-25.44~  & ~~~~108.5~~~~ & ~~~~7.014~~~~ &   ~-165.0~    & ~5.619~&    ~18.40~\\
\hline\hline
\end{tabular}
\end{center}
\end{table}
From this table, it is shown that the major attraction for binding comes from the $\sigma$-meson exchange, and the main repulsion for unbinding comes from the $\omega$-meson exchange and the kinetic energy. The averaged contributions from the $\pi$-meson exchange, which also shows the repulsive character, and from the $\rho$-meson exchanges, whose character depends on the case, are much smaller than those from the $\sigma$- and $\omega$-exchanges.

\begin{figure}[htbp]
  \begin{center}
  \rotatebox{0}{\includegraphics*[width=0.47\textwidth]{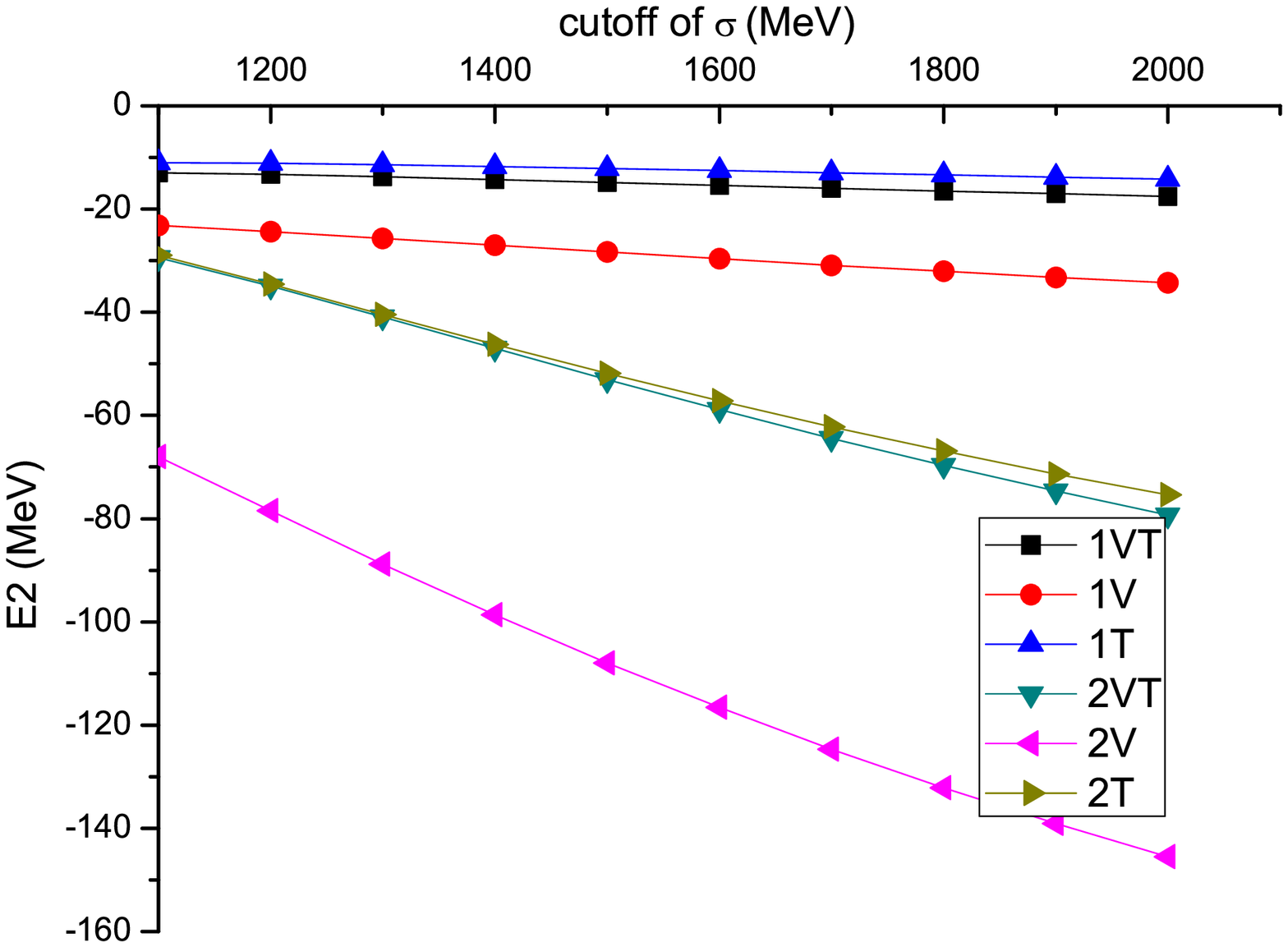}}
    \rotatebox{0}{\includegraphics*[width=0.47\textwidth]{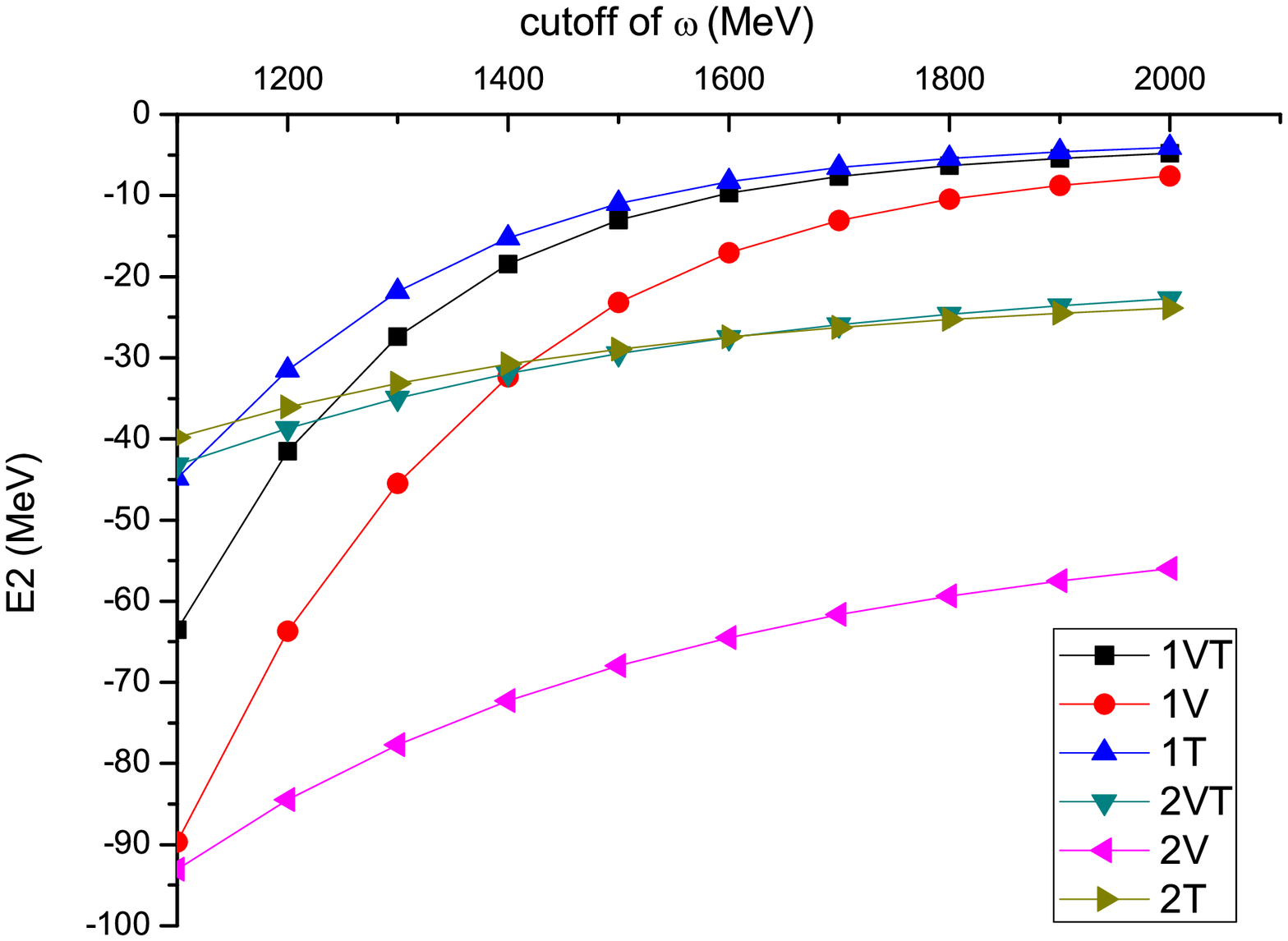}}
      \rotatebox{0}{\includegraphics*[width=0.47\textwidth]{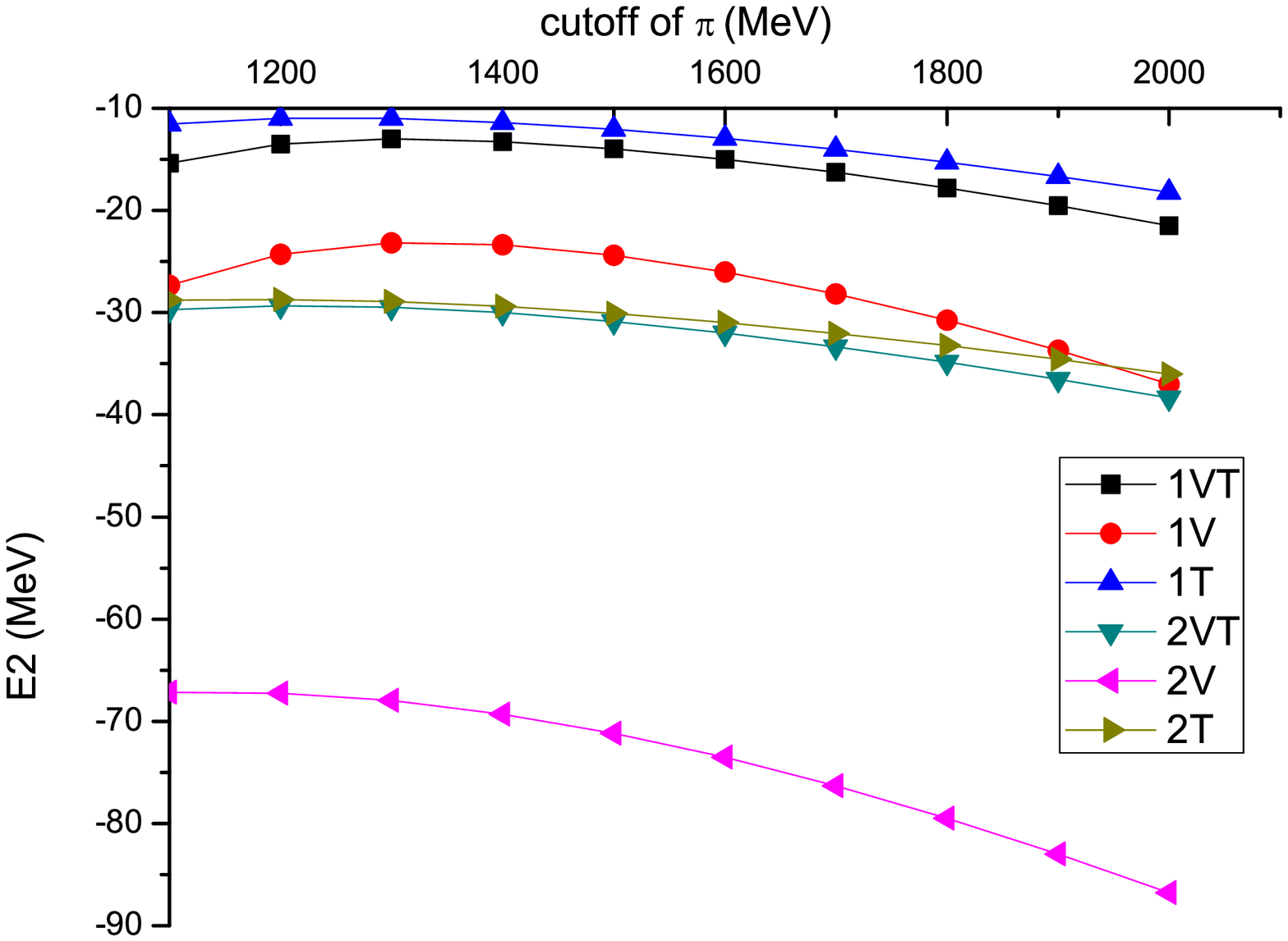}}
        \rotatebox{0}{\includegraphics*[width=0.47\textwidth]{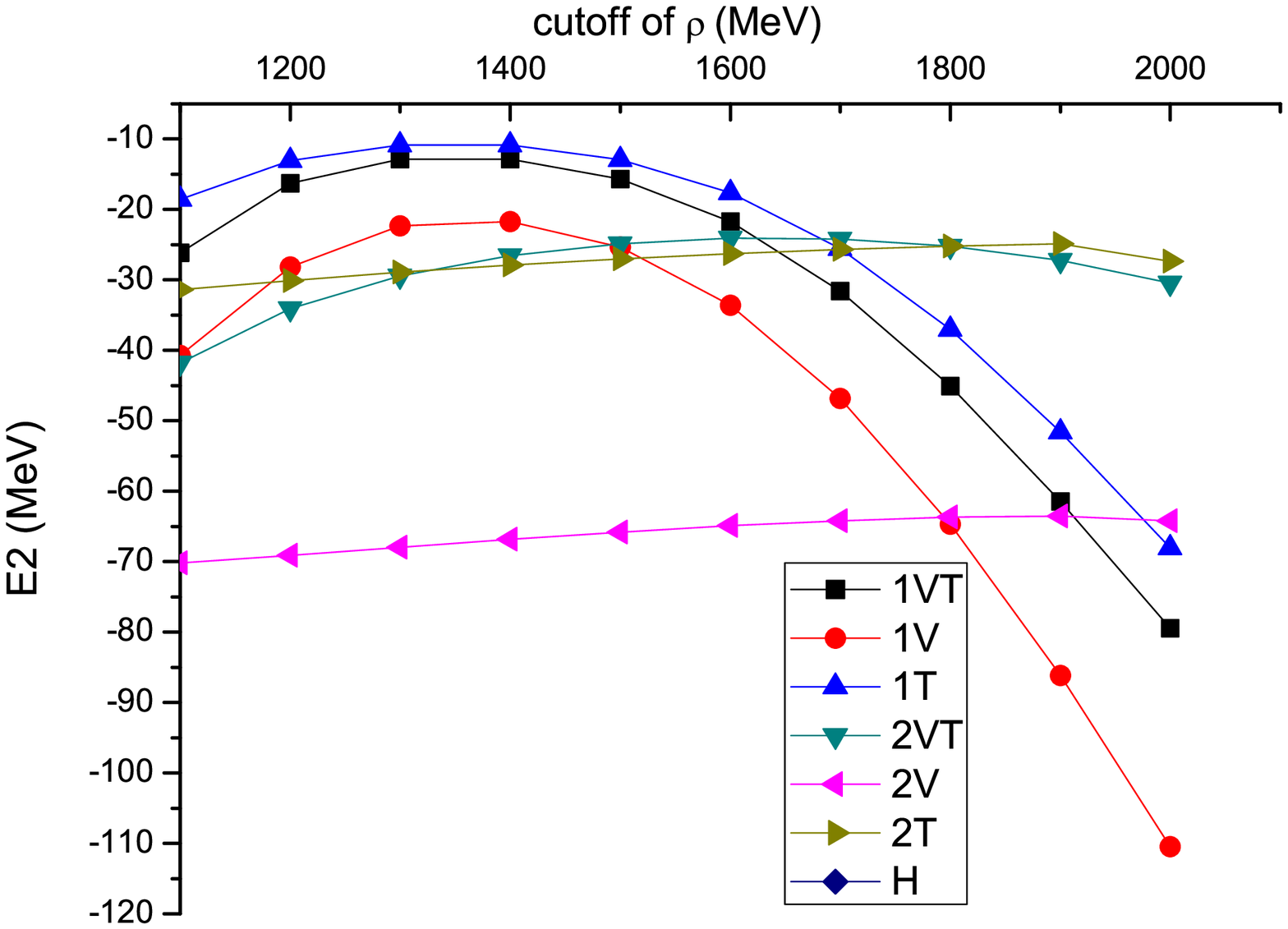}}
    \caption{ Binding energy $E_2$ as a function of $\Lambda_{\alpha}$ ($\alpha=\sigma$(upper left graph), $\omega$(upper right graph), $\pi$(lower left graph) and $\rho$(lower left graph)). }
    \label{fig:lambda}
  \end{center}
\end{figure}

As mentioned previously, before realistically studying the binding property of the $NN^*$(1440) system, the cutoff values should be carefully re-adjusted due to the larger size of $N^*$. In other word, the obtained result should be stable with respect to the model parameters, namely the deviations of the cutoff parameter values, thus the potential model would have prediction power. Here, we plot the binding energy $E_2$ as the function of $\Lambda_{\alpha}$ ($\alpha=\sigma$(upper left graph), $\omega$(upper right graph), $\pi$(lower left graph) and $\rho$(lower left graph)) in Fig.~\ref{fig:lambda}. The curves for $E_1$ are very similar, so we will not present them here.
From this figure, one sees again the above mentioned features: (1). The binding energy in the single tensor coupling case is close to that in the vector+tensor coupling case, which implies that we can use a single tensor coupling mode instead of the vector+tensor coupling mode for simplicity. (2). The binding energy in the single vector coupling case is larger than that in the single T or V+T cases in the Case 2. (3). The binding energy in the Case 2 is larger than that in the Case 1 due to more attractive feature of the potential in the Case 2 than in the Case 1. Besides, there exist different features for various meson exchanges. For the $\sigma$-exchange (upper left diagram in Fig.~\ref{fig:lambda}), we have following observations: (1). The binding energy increases with increasing $\Lambda_{\sigma}$ because of an attractive feature of the $\sigma$-exchange. (2). The result in the Case 1 is stable along with the variation of $\Lambda_{\sigma}$. But, in the Case 2, the result increases rapidly with increasing $\Lambda_{\sigma}$. This is because that the averaged contribution from the $\sigma$-exchange is much larger than those from the other exchanges, thus the inconsistency between the coupling constants of a symmetric $N^*$-$\sigma$-$N^*$ interaction and those of an asymmetric $N$-$\sigma$-$N^*$ interaction would be enlarged. It also implies that the results in the Case 1 might be more meaningful. For the $\omega$-exchange (upper right diagram in Fig.~\ref{fig:lambda}), it shows: (1). The binding energy decreases with increasing $\Lambda_{\omega}$ because of a repulsive feature of the $\omega$-exchange. (2). The result is stable as long as $\Lambda_{\omega}\geq 1.8GeV$. (3). When $\Lambda_{\omega}\leq 1.2GeV$, the result would be unstable, because too much repulsion is cut due to a smaller $\Lambda_{\omega}$ value. For the $\pi$-exchange (lower left diagram in Fig.~\ref{fig:lambda}), we see: (1). The binding energy increases with increasing $\Lambda_{\pi}$ in the smaller $\Lambda_{\pi}$ region. (2). The result is relative stable as long as $\Lambda_{\pi}\simeq 1.1 \sim 1.3GeV$. For the $\rho$-exchange (lower right graph of Fig.~\ref{fig:lambda}), we get: (1). The binding energy firstly increases then decreases with increasing $\Lambda_{\rho}$, because the character of the average contribution from $\rho$-exchange depends on how much repulsion being cut. (2). The result is stable when $\Lambda_{\rho}\simeq 1.2 \sim 1.5GeV$ only. Form above analysis, we find that in order to have a meaningful result, we should take the values of $\Lambda_{\alpha}$ within the following ranges: $\Lambda_{\pi}\simeq 1.1 \sim 1.3GeV$, $\Lambda_{\sigma}\geq 1.1GeV$, $\Lambda_{\rho}\simeq 1.2 \sim 1.5GeV$ and $\Lambda_{\omega}\geq 1.8GeV$. Therefore, in the rest of the calculation, we take $\Lambda_{\pi}= 1.1GeV$, $\Lambda_{\sigma}= 1.2GeV$, $\Lambda_{\rho}= 1.3GeV$ and $\Lambda_{\omega}= 1.8GeV$, respectively. With this set of cutoff parameters, called Set $B_{\Lambda}$, we re-calculate all the coupling constants and tabulate them in Table~\ref{tab:couplingB}. The corresponding binding energies are tabulated in Table~\ref{tab:cen bound states B set}.
\begin{table}[htbp]
\centering \caption{$g_{\alpha NN^*(1440)}$, $f_{\alpha NN^*(1440)}$ and $\Lambda_\alpha$ in Set $B_{\Lambda}$.}
 \label{tab:couplingB}
\begin{center}
\begin{tabular}{c c c c c c}
\hline\hline
\multirow{2}{*}{Coupling mode of $\rho$~~~~~} & \multirow{2}{*}{~~~Coupling constant~~~~~~}  &
\multicolumn{4}{c}{$\alpha$}  \\
\cline{3-6}
   &   & $~~~~~~\pi~~~~~~$ & $~~~~~~\sigma~~~~~~$  & $~~~~~~\rho~~~~~~$ & $~~~~~~\omega~~~~~~$ \\
\hline
\multirow{2}{*}{V+T} & $g^{2}_{\alpha NN^*(1440)}/{4\pi}$ & 2.9087  & 4.900             &  0.3685 & 1.1187\\
\cline{2-6}
 & $f^{2}_{\alpha NN^*(1440)}/{4\pi}$ & --- & --- &  2.8288 & ---\\
\hline
\multirow{2}{*}{V} & $g^{2}_{\alpha NN^*(1440)}/{4\pi}$ & 2.9087    & 4.900             &  1.1374 & 1.1187\\
\cline{2-6}
 & $f^{2}_{\alpha NN^*(1440)}/{4\pi}$ & --- & --- &  --- & ---\\
\hline
\multirow{2}{*}{T} & $g^{2}_{\alpha NN^*(1440)}/{4\pi}$ & 2.9087    & 4.900             &  --- & 1.11870\\
\cline{2-6}
 & $f^{2}_{\alpha NN^*(1440)}/{4\pi}$ & --- & --- &  1.1374 & ---\\
\hline\hline
\multirow{2}{*}{V+T} & $g^{2}_{\alpha NN}/{4\pi}$ & 14.900  & 7.7823 &  0.9500 & 20.000\\
\cline{2-6}
 & $f^{2}_{\alpha NN}/{4\pi}$ & --- & --- &  35.350 & ---\\
\hline
 & $\Lambda_\alpha$ (GeV) & 1.1000 & 1.2000 &  1.3000 & 1.8000\\
\hline
\hline
\end{tabular}
\end{center}
\end{table}
\begin{table}[htbp]
\caption{Binding energies of the $NN^*$(1440) system (in MeV) in Set $B_{\Lambda}$. }
\label{tab:cen bound states B set}
\begin{center}
\begin{tabular}{c c c c}
\hline
\hline
\multirow{2}{*}{Coupling mode of $\rho$~~~~~} & \multirow{2}{*}{~~~~~~~~}  &
\multicolumn{2}{c}{Binding energy}  \\
\cline{2-4}
   &   &~~~~~~~~ case 1~~~~~~~~ &~~~~~~~~ case 2~~~~~~~~  \\
\hline
\multirow{2}{*}{V} & $E_1$ &   ~-296.5~      &  ~-596.6~\\
\cline{2-4}
                    & $E_2$ &   ~-10.2~       &  ~-68.1~\\
\hline
\multirow{2}{*}{T} & $E_1$ &   ~-253.0~      &  ~-440.1~\\
\cline{2-4}
                   & $E_2$ &   ~-4.7~        &  ~-30.0~\\
\hline
\multirow{2}{*}{V+T} & $E_1$ &  ~-261.2~      &  ~-436.4~\\
\cline{2-4}
                     & $E_2$ & ~-5.7~       &  ~-31.2~ \\
\hline\hline
\end{tabular}
\end{center}
\end{table}
From this table, we again see that the binding energies in the case with a single vector coupling mode are usually larger than those in the case with a single tensor  coupling mode or a vector+tensor coupling mode, and the binding energies in the later two cases are comparable. For self-consistency, we would ignore the single V coupling mode in the later calculation. Moreover, the results in the Case 2 are generally larger than those in the Case 1. All these observations agree with the potentials shown in the previous section.


We also present the variational ranges of coupling constants with Set $B_{\Lambda}$ due to the uncertainties of the decay data for $N^*$(1440) in Table~\ref{tab:couplingC} and corresponding ranges of binding energies in Table~\ref{tab:var bound states B set}.
\begin{table}
\caption{Variational ranges of coupling constants in Set $B_{\Lambda}$.} \label{tab:couplingC}
\begin{footnotesize}
\begin{center}
\begin{tabular}{c c c c c c}
\hline\hline
\multicolumn{2}{c}{~~} & $~~~~~~~~~\pi~~~~~~~~~~~$ & $~~~~~~~~~~~\sigma~~~~~~~~~$  & $~~~~~~~~~\rho~~~~~~~~~$ & $~~~~~~~~~\omega~~~~~~~~~$ \\
\hline
\multirow{2}{*}{Branching ratio}& &
  $~~~~\Gamma_{\pi}/\Gamma~~~~$ & $~~~~\Gamma_{\sigma}/\Gamma~~~~$  & $~~~~\Gamma_{N^*\to N\gamma~(I_{\gamma}=1)}~~~~$  &  $\Gamma_{N^*\to N\gamma~(I_{\gamma}=0)}$\\
\cline{2-6}
                              & & $0.55\sim 0.75$ & $0.11\sim 0.13$ & $0.074\sim 0.13$ &  $0.0011\sim 0.014$\\
\hline\hline
$g^{2}_{\alpha NN^*(1440)}/{4\pi}$ & & $2.461\sim 3.356~~$ & $~~4.492\sim 5.308$~~~    & ---
& ~~~ $0.2166\sim 2.722$\\
\hline
$f^{2}_{\alpha NN^*(1440)}/{4\pi}$ & & --- & ---    & $0.8543\sim 1.461$ (T case)&  ---\\
\hline\hline
$g^{2}_{\alpha NN^*(1440)}/{4\pi}$ & &$2.461\sim 3.356~~$ & $~~4.492\sim 5.308$~~~    & $0.02282\sim 0.9717$(VT case) &~~~ $0.2166\sim 2.722$\\
\hline
$f^{2}_{\alpha NN^*(1440)}/{4\pi}$ & & --- & ---    & $3.779\sim 2.084$(VT case)&  ---\\
\hline\hline
$\Lambda_{\alpha}$(GeV) &        & 1.1  & 1.2   & 1.3 & 1.8 \\
\hline\hline
\end{tabular}
\end{center}
\end{footnotesize}
\end{table}
\begin{table}[htbp]
\caption{Binding energies of the $NN^*$(1440) system (in MeV) in Set $B_{\Lambda}$. }
\label{tab:var bound states B set}
\begin{center}
\begin{tabular}{c c c c}
\hline
\hline
\multirow{2}{*}{Coupling mode of $\rho$~~~~~} & \multirow{2}{*}{~~~~~~~~}  &
\multicolumn{2}{c}{Binding energy}  \\
\cline{2-4}
   &   &~~~~~~~~ case 1~~~~~~~~ &~~~~~~~~ case 2~~~~~~~~  \\
\hline\hline
\multirow{2}{*}{T} & $E_1$ &   $-222.3\!\sim\!-285.0$      &  $-324.2\!\sim\!-584.5$\\
\cline{2-4}
                   & $E_2$ &   $-2.34\!\sim \!-8.330$        & $-12.16\!\sim\!-62.61$\\
\hline\hline
\multirow{2}{*}{V+T} & $E_1$ &  $-230.2\!\sim\!-293.7$      &  $-330.9\!\sim\!-562.6$\\
\cline{2-4}
                     & $E_2$ & $-2.897\!\sim \!-9.987$       &  $-13.38 \!\sim\!-67.43$ \\
\hline\hline
\end{tabular}
\end{center}
\end{table}
From this table, one sees that because of many uncertain factors, for instance the large uncertainties of the $N^*(1440)$ decay data, the indeterminate coupling mode of the $\rho$-meson, the coupling constants of $g_{N^*N^*M}$ and $f_{N^*N^*M}$ and etc., the resultant binding energies spread in the large regions, say about $222\sim 584$MeV for the ground state and about $2\sim 67$MeV for the first excited state.

Before closing this section, we would mention that in the $N$-$N$ scattering, nucleon might be excited to $N^*$. So, the deuteron could be a mixture of the $S-$wave ground state with other high partial wave states, say the $D$-wave state, and even $NN^*$ or $N^*N^*$ states, as long as the quantum numbers are allowed. Because the deeply bound state of $NN^*$(1440) with a binding energy of about a few hundred MeV is close to the energy level of the deuteron, it might couple to the conventional deuteron to form a realistic deuteron. Namely, this state would be a component of the real deuteron. Therefore, one may not be easy to observe it in experiment. However, the second bound state just has a binding energy of about a few tens of MeV, say $2 MeV\sim 67 MeV$ in this calculation, it is possible to be observed in experiment. In fact, this state is very close to the newly observed resonance-like structure around 2360 MeV in the $pn\to d\pi^0\pi^0$ reaction cross section data. Therefore, we speculate that the observed state may have a $NN^*$(1440) structure.

\section{Summary}

The meson-exchange model is extended to study the $NN^*(1440)$ interaction. In this model, the $\pi$-, $\sigma$-, $\rho$- and $\omega$-meson exchanges between $N$ and $N^*$(1440) are assumed. By calculating $t$- and $u$-channel Feynman diagrams, the $N$-$N^*(1440)$ potential is derived. The coupling constants $g_{\alpha NN}^2/4\pi$ and $f_{\alpha NN}^2/4\pi$ are taken from the Bonn potential model. The coupling constants $g_{\alpha NN^*(1440)}^2/4\pi$ and $f_{\alpha NN^*(1440)}^2/4\pi$ are phenomenologically determined by fitting the partial decay width data of the $N^*\to N\pi$, $N^*\to N\pi\pi$, $N^*\to N\rho \to N\pi\pi$ and $N^*\to N\gamma$ processes. As to the coupling constants $g_{\alpha N^*(1440)N^*(1440)}^2/4\pi$ and $f_{\alpha N^*(1440)N^*(1440)}^2/4\pi$, two cases are assumed. It is found that similar to the $N$-$N$ potential, the $\sigma$-meson exchange provides a major attraction, the $\omega$-meson exchanges mainly supplies a repulsive force, $\pi$- and $\rho$-meson exchanges give an effective medium- and long-range attractions. Solving the Schr\"{o}dinger equation by a matrix method, the binding energies and corresponding wave functions are obtained. It is shown that there might exist two $N$-$N^*$(1440) $S$-wave bound states. One of them is deeply bound with a binding energy ranged from 222MeV to 584 MeV. This state could be a small component of the real deuteron. Another state is weakly bound with a binding energy ranged from 2MeV to 67 MeV. This state is very close to the newly observed resonance-like structure around 2360 MeV in the $pn\to d\pi^0\pi^0$ reaction cross section. Therefore, we speculate that the observed state may have a large component of the $NN^*$(1440) quasi-molecular structure.

\section{Appendix}

\begin{eqnarray}
\mathcal{F}_{1t}=&&\mathcal{F}\{(\frac{\Lambda^2-m^2}{{\Lambda}^2+\vec{q}^2})
\frac{1}{\vec{q}^2+m^2}\}\nonumber\\
=&&mY(mr)-\Lambda Y(\Lambda r)- (\Lambda^2-m^2)\frac{e^{-\Lambda r}}{2\Lambda}
\end{eqnarray}

\begin{eqnarray}
\mathcal{F}_{1u}=&&\mathcal{F}\{(\frac{\Lambda^2-m^2}{\tilde{\Lambda}^2+\vec{q}^2})
\frac{1}{\vec{q}^2+M^2}\}\nonumber\\
=&&MY(Mr)-\tilde{\Lambda} Y(\tilde{\Lambda} r)-(\Lambda^2-m^2)\frac{e^{-\tilde{\Lambda} r}}{2\tilde{\Lambda}}
\end{eqnarray}

\begin{eqnarray}
\mathcal{F}_{2t}=&&\mathcal{F}\{(\frac{\Lambda^2-m^2}{{\Lambda}^2+\vec{q}^2})
\frac{\vec{q}^2}{\vec{q}^2+m^2}\}\nonumber\\
=&&m^2[~\Lambda Y(\Lambda r)-mY(mr)~]+(\Lambda^2-m^2)\Lambda\frac{e^{-\Lambda r}}{2}
\end{eqnarray}

\begin{eqnarray}
\mathcal{F}_{2u}=&&\mathcal{F}\{(\frac{\Lambda^2-m^2}{{\tilde{\Lambda}}^2+\vec{q}^2})
\frac{\vec{q}^2}{\vec{q}^2+M^2}\}\nonumber\\
=&&M^2[~\tilde{\Lambda} Y(\tilde{\Lambda} r)-MY(Mr)~]+(\Lambda^2-m^2)\tilde{\Lambda}\frac{e^{-\tilde{\Lambda} r}}{2}
\end{eqnarray}

\begin{eqnarray}
\mathcal{F}_{3t} = && \mathcal{F}\{(\frac{\Lambda^2-m^2}{{\Lambda}^2+\vec{q}^2})
\frac{(\vec{\sigma_1}\cdot\vec{q})(\vec{\sigma_2}\cdot\vec{q})}{\vec{p}^2+m^2}\}\nonumber\\
=&&\frac{1}{3}\vec{\sigma_1}\cdot\vec{\sigma_2}[~m^2\Lambda Y(\Lambda r)-m^3 Y(mr)
+(\Lambda^2-m^2)\Lambda\frac{e^{-\Lambda r}}{2}~]\nonumber\\
&&+\frac{1}{3}S_{12}[-m^3 Z(mr)+ \Lambda^3 Z(\Lambda r) + (\Lambda^2-m^2)(1+\Lambda r)
\frac{\Lambda}{2}Y(\Lambda r)~]\nonumber\\
=&&(\vec{\sigma_1}\cdot\vec{\sigma_2})\mathcal{F}_{3t1} + S_{12}\mathcal{F}_{3t2}
\end{eqnarray}

\begin{eqnarray}
\mathcal{F}_{3u} = && \mathcal{F}\{(\frac{\Lambda^2-m^2}{{\tilde{\Lambda}}^2+\vec{q}^2})
\frac{(\vec{\sigma_1}\cdot\vec{q})(\vec{\sigma_2}\cdot\vec{q})}{\vec{q}^2+M^2}\}\nonumber\\
=&&\frac{1}{3}\vec{\sigma_1}\cdot\vec{\sigma_2}[~M^2\tilde{\Lambda} Y(\tilde{\Lambda} r)-M^3Y(Mr)
+(\Lambda^2-m^2)\tilde{\Lambda}\frac{e^{-\tilde{\Lambda} r}}{2}~]\nonumber\\
&&+\frac{1}{3}S_{12}[-M^3 Z(Mr)+ \tilde{\Lambda}^3 Z(\tilde{\Lambda} r) + (\Lambda^2-m^2)
(1+\tilde{\Lambda} r)\frac{\tilde{\Lambda}}{2}Y(\tilde{\Lambda} r)~]\nonumber\\
=&&(\vec{\sigma_1}\cdot\vec{\sigma_2})\mathcal{F}_{3u1} + S_{12}\mathcal{F}_{3u2}
\end{eqnarray}

\begin{eqnarray}
\mathcal{F}_{4t} =&&\mathcal{F}\{{(\frac{\Lambda^2-m^2}{{\Lambda}^2+\vec{q}^2})
\frac{\vec{k}^2}{\vec{q}^2+m^2}}\}\nonumber\\
=&&\frac{m^3}{4}Y(mr)-\frac{\Lambda^3}{4}Y(\Lambda r)-\frac{\Lambda^2-m^2}{4}(\frac{\Lambda r}{2}-1)
\frac{e^{-\Lambda r}}{r}\nonumber\\
&&-\frac{1}{2}\{\nabla^2,m Y(mr)-\Lambda Y(\Lambda r)-\frac{\Lambda^2-m^2}{2}\frac{e^{-\Lambda r}}{\Lambda}\}\nonumber\\
=&&\mathcal{F}_{4t1}+\{-\frac{1}{2}\nabla^2,\mathcal{F}_{4t2}\}
\end{eqnarray}

\begin{eqnarray}
\mathcal{F}_{4u} =&&\mathcal{F}\{{(\frac{\Lambda^2-M^2}{{\tilde{\Lambda}}^2+\vec{q}^2})
\frac{\vec{k}^2}{\vec{q}^2+M^2}}\}\nonumber\\
=&&\frac{M^3}{4}Y(Mr)-\frac{\tilde{\Lambda}^3}{4}Y(\tilde{\Lambda} r)-
\frac{\Lambda^2-m^2}{4}(\frac{\tilde{\Lambda} r}{2}-1)\frac{e^{-\tilde{\Lambda} r}}{r}\nonumber\\
&&-\frac{1}{2}\{\nabla^2,M Y(Mr)-\tilde{\Lambda} Y(\tilde{\Lambda} r)-\frac{\Lambda^2-m^2}{2}
\frac{e^{-\tilde{\Lambda} r}}{\tilde{\Lambda}}\}\nonumber\\
=&&\mathcal{F}_{4u1}+\{-\frac{1}{2}\nabla^2,\mathcal{F}_{4u2}\}
\end{eqnarray}

\begin{eqnarray}
\mathcal{F}_{5t} =&&\mathcal{F}\{{i(\frac{\Lambda^2-m^2}{{\Lambda}^2+\vec{q}^2})
\frac{\vec{S}\cdot(\vec{q}\times\vec{k})}{\vec{q}^2+m^2}}\}\nonumber\\
=&&\vec{S}\cdot\vec{L}[~-m^3 Z_{1}(mr)+\Lambda^3 Z_{1}(\Lambda r)+(\Lambda^2-m^2)
\frac{e^{-\Lambda r}}{2r}~]\nonumber\\
=&&\vec{S}\cdot\vec{L}\mathcal{F}_{5t0}
\end{eqnarray}

\begin{eqnarray}
\mathcal{F}_{5u} =&&\mathcal{F}\{{i(\frac{\Lambda^2-m^2}{{\tilde{\Lambda}}^2+\vec{q}^2})
\frac{\vec{S}\cdot(\vec{q}\times\vec{k})}{\vec{q}^2+M^2}}\}\nonumber\\
=&&\vec{S}\cdot\vec{L}[~-M^3 Z_{1}(Mr)+\tilde{\Lambda}^3 Z_{1}(\tilde{\Lambda} r)+(\Lambda^2-m^2)
\frac{e^{-\tilde{\Lambda} r}}{2r}~]\nonumber\\
=&&\vec{S}\cdot\vec{L}\mathcal{F}_{5u0}
\end{eqnarray}

\begin{eqnarray}
\mathcal{F}_{6u}= && \mathcal{F}\{(\frac{\Lambda^2-m^2}{{\tilde{\Lambda}}^2+\vec{q}^2})
\frac{(\vec{\sigma_1}\cdot\vec{q})(\vec{\sigma_2}\cdot\vec{q})}{\vec{p}^2-M^2}\}\nonumber\\
=&&\frac{1}{3}\vec{\sigma_1}\cdot\vec{\sigma_2}[~-M^2\tilde{\Lambda} Y(\tilde{\Lambda} r)-M^3\frac{\cos(Mr)}{Mr}
+(\Lambda^2-m^2)\tilde{\Lambda}\frac{e^{-\tilde{\Lambda} r}}{2}~]\nonumber\\
&&+\frac{1}{3}S_{12}[M^3 Z'(Mr)+ \tilde{\Lambda}^3 Z(\tilde{\Lambda} r) + (\Lambda^2-m^2)
(1+\tilde{\Lambda} r)\frac{\tilde{\Lambda}}{2}Y(\tilde{\Lambda} r)~]\nonumber\\
=&&(\vec{\sigma_1}\cdot\vec{\sigma_2})\mathcal{F}_{6u1} + S_{12}\mathcal{F}_{6u2}
\end{eqnarray}

\begin{acknowledgments}

We would like to thank X.Cao, J.J. Wu, and J.J. Xie for their
helpful discussions. This work is supported in part by the National
Natural Science Foundation of China under Grant 10975038, 11035006,
11165005, 11121092, 11261130311 (CRC110 by DFG and NSFC), the
Chinese Academy of Sciences under Project No.KJCX2-EW-N01 and the
Ministry of Science and Technology of China (2009CB825200).

\end{acknowledgments}

\end{document}